\documentclass[11pt,a4paper]{article}
\pdfoutput=1
\usepackage{jcappub}
\usepackage{multirow}
\usepackage{array}
\newcommand{\comment}[1]{}

\newcommand{\Rvir}{R_\mathrm{vir}}
\newcommand{\Mvir}{M_\mathrm{vir}}

\title{Unbound Particles in Dark Matter Halos}

\author[a]{Peter S. Behroozi,}
\author[b]{Abraham Loeb,}
\author[a]{Risa H. Wechsler}
\affiliation[a]{Kavli Institute for Particle Astrophysics and
  Cosmology; Physics Department, Stanford University, and SLAC National
  Accelerator Laboratory\\ 2575 Sand Hill Road, Menlo Park, CA, USA}
\affiliation[b]{Department of Astronomy, Harvard University\\ 60 Garden St, Cambridge, MA, USA}
\emailAdd{behroozi@stanford.edu}
\abstract{We investigate unbound dark matter particles in halos by tracing particle trajectories in a simulation run to the far future ($a=100$).  We find that the traditional sum of kinetic and potential energies is a very poor predictor of which dark matter particles will eventually become unbound from halos.  We also study the mass fraction of unbound particles, which increases strongly towards the edges of halos, and decreases significantly at higher redshifts.  We discuss implications for dark matter detection experiments, precision calibrations of the halo mass function, the use of baryon fractions to constrain dark energy, and searches for intergalactic supernovae.}

\keywords{dark matter, N-body simulations, cosmology:theory}

\begin{document}
\maketitle
\flushbottom

\newcommand{\Msun}{M_{\odot}}
\newcommand{\mvir}{M_\mathrm{vir}}
\newcommand{\rvir}{R_\mathrm{vir}}
\newcommand{\vmax}{v_\mathrm{max}}
\newcommand{\plotgrace}[1]{\includegraphics[width=0.5\columnwidth,type=pdf,ext=.pdf
,read=.pdf]{#1}}
\newcommand{\plotlargegrace}[1]{\includegraphics[width=0.7\columnwidth,type=pdf,ext=.pdf
,read=.pdf]{#1}}
\newcommand{\plotminigrace}[1]{\includegraphics[width=0.5\columnwidth,type=pdf,ext=.pdf
,read=.pdf]{#1}}

\section{Introduction}

In an expanding cosmology, a particle's trajectory depends not only on its initial velocity, but also on the dynamical evolution of nearby structure.  The meaning of ``unbound'' in this case is both nontrivial and physically interesting.  Traditionally, this word describes particles which can escape to arbitrarily large distances from their initial host dark matter halo.  We investigate the basic demographics of unbound dark matter by tracing particle trajectories in cosmological simulations.  In addition, we also investigate traditional unbound particle classification techniques based on kinetic energies.

While intellectual curiosity motivated this work, there are many practical implications.  For example, we find that the fractional abundance of high-energy dark matter particles depends strongly on halo environment; e.g., the distance to the nearest larger halo.  This suggests that the velocity distribution of dark matter depends on halo environment, which is important for dark matter direct detection experiments \citep{XENON100,CRESST,DAMA,COGENT,CDMS,Mao12,Lisanti11,PS1,PS2,PS3,PS4,PS5,PS6,PS7}.  We therefore test the expected effect from Andromeda on the Milky Way's dark matter velocity distribution directly.

Dark matter particle trajectories also have an effect on the baryon fraction in halos.  When a high-energy dark matter particle leaves the virial radius, the virial mass of the halo is lowered.  The hot halo gas will expand adiabatically in the reduced potential, nominally leaving the hot gas fraction unchanged.  However, the centrally-concentrated stars will not be able to expand, meaning that the overall baryon fraction will increase.  This effect will result in a redshift- and halo-mass-dependent baryon fraction.  Of course, the true picture is made more complicated by a number of effects, including angular momentum exchange, gas cooling, and galaxy feedback.  Nonetheless, it is possible to make a straightforward estimate of its magnitude. In the future, constraining this effect may be important for surveys which aim to place precision constraints on cosmology (e.g., BOSS, DES, BigBOSS, Pan-STARRS, eRosita, Herschel, Planck, JWST, and LSST; \cite{BOSS,DES,BigBOSS,Pan-STARRS,eROSITA,Herschel,Planck,JWST,LSST}).  In order to fully realize their statistical power, these surveys depend on precision calibrations of the halo mass function to 1\% or better accuracy \cite{Wu10,Cunha10}.

Observations of intergalactic supernovae \citep{Sharon10,Sand11,Barbary12} relate to unbound particles, especially observations of supernovae traveling at high speeds \citep{GalYam03} and hostless gamma-ray bursts (GRBs) \cite{Fong13}.  Such escaping objects might either come from the intracluster stellar population \citep{Monaco06,Barbary12}, from merging events \citep{Murante07,Teyssier09}, or from hypervelocity stars ejected from the central galaxy \citep{Napiwotzki11}.  Since stars are effectively collisionless and so evolve dynamically like dark matter particles, one can place upper limits on the fraction of supernova and GRB progenitors ejected through merger events.

In \S \ref{s:methods}, we discuss the dark matter simulations we employ as well as the halo finder and conventions used for calculating kinetic and potential energies.  We discuss unique qualitative features of unbound particles in \S \ref{s:energy} and derive kinetic escape thresholds; we also test how well kinetic escape thresholds can classify unbound particles in full cosmological simulations.  We present the main results for the population of unbound particles in \S \ref{s:results} and discuss how these results impact the science considerations above in \S \ref{s:discussion}.  Finally, we summarize our conclusions in \S \ref{s:conclusions}.  We use simulations with multiple cosmologies in this work, but our primary results assume a flat, $\Lambda$CDM cosmology with main parameters $\Omega_b = 0.04$, $\Omega_M = 0.27$, $\Omega_\Lambda = 0.73$, and $h = 0.7$.  All halo masses at $a\le1$ are defined using the virial overdensity criterion \cite{mvir_conv}.

\section{Methods and Conventions}

\label{s:methods}

\subsection{Simulation}

The primary simulation used is a large (1 Gpc $h^{-1}$ on a side) collisionless dark matter volume run from early redshifts ($a=0.02$) to the far future ($a=100$) using \textsc{gadget}-2 \citep{Springel05}.  This simulation traces $1024^3$ particles, with a particle mass resolution of $10^{11}\Msun$ and a spline force softening of $30$ kpc $h^{-1}$.  This makes it ideal for particle studies from group-scale halos ($10^{13.6}\Msun$, 400 particles) to massive clusters ($10^{15}\Msun$, 10000 particles).  The initial conditions were generated from a flat, $\Lambda$CDM cosmology ($\Omega_M = 0.27$, $\Omega_\Lambda = 0.73$, $h = 0.7$, $\sigma_8 = 0.79$, and $n_s = 0.95$), which is similar to the WMAP7 best-fit cosmology \citep{wmap7}.  From this simulation, 300 snapshots were saved from $a=0.075$ to $a=100$, spaced at uniform logarithmic intervals in scale factor.

We also make use of the \textit{Consuelo} simulation from the Large Suite of Dark Matter Simulations (McBride et al., in preparation).\footnote{LasDamas Project, {\tt http://lss.phy.vanderbilt.edu/lasdamas/}} Consuelo covers a large volume (420 $h^{-1}$ Mpc on a side) with $\sim$2.7 billion particles ($1400^3$), corresponding to a particle mass resolution of $2.7 \times 10^9$ $\Msun$.  Its large size and resolution make it ideal for more detailed particle studies from Milky Way-sized halos ($10^{12}\Msun$, 400 particles) to the largest clusters ($10^{15}\Msun$, 400,000 particles).  This simulation was also performed using \textsc{gadget}-2, with a spline force softening of 8 $h^{-1}$ kpc.  The assumed simulation conditions were a flat, $\Lambda$CDM cosmology ($\Omega_M = 0.25$, $\Omega_\Lambda = 0.75$, $h = 0.7$, $\sigma_8 = 0.8$, and $n_s = 1.0$); 100 snapshots were saved from $a=0.075$ to $a=1.0$, also spaced at uniform logarithmic intervals in scale factor.

In a few cases where resolution tests are necessary, we make use of the \textit{Esmeralda} simulation, also from the LasDamas Project.  This simulation was run with identical parameters and software as the Consuelo simulation, with the exception of its box size (640 $h^{-1}$ Mpc, particle count ($1250^3$), mass resolution ($1.33 \times 10^{10}$ $\Msun$, and spline force softening (15 h$^{-1}$ kpc).  For one test, we also make use of the \textit{Bolshoi} simulation \cite{Bolshoi}, which has much higher resolution ($2048^3$ particles, each $1.9 \times 10^8 \Msun$, with force resolution of 1 $h^{-1}$ kpc) but a much smaller volume of 250 $(h^{-1}$ Mpc$)^3$.  Bolshoi was run with the \textsc{art} code \cite{kravtsov_etal:97}, and it assumed a flat, $\Lambda$CDM cosmology similar to our main simulation ($\Omega_M = 0.27$, $\Omega_\Lambda = 0.73$, $h = 0.7$, $\sigma_8 = 0.82$, and $n_s = 0.96$).

\subsection{Halo Finding and Properties}

\label{s:halo_finding}

Halo finding was performed using the \textsc{rockstar} algorithm, which is a phase-space temporal (7D) halo finder designed for high consistency and accuracy of halo properties \citep{Rockstar,BehrooziTree}.  The \textsc{rockstar} halo finder locates peaks in the particle phase-space density distribution using a locally adaptive phase-space metric.  A seed halo is placed at the location of each peak, and particles are assigned to the closest seed halo in phase space (see \cite{Rockstar} for full details).  Particle potentials are computed using a tree method (see \S\ref{s:grav_pot}); contrary to the default setting, we do not remove positive-energy particles for the analysis in this paper.  For cases where host/satellite halo relationships are ambiguous (such as in major mergers), \textsc{rockstar} uses the host/satellite relationship at the previous snapshot if available.

In this analysis, we consider host halos only (i.e., halos whose centers are not within the radius of more massive halos).  This is because satellite halos (i.e., non-host halos) are almost always \textit{defined} as a collection of self-bound particles within a larger halo---meaning that unbound particles are automatically excluded.  Moreover, there are many reasonable choices of satellite halo boundaries \cite{Onions12}, which makes essential quantities like gravitational potentials depend strongly on the halo finder used.

We also restrict our analysis in general to halos with more than 400 particles, which have been shown to have robustly measurable properties when compared to higher-resolution resimulations \citep{Trenti10}.  We have performed resolution tests with the Esmeralda simulation to verify that gravitational potential calculations for 400-particle halos result in identical energy distributions as for the equivalent 2000-particle halos in Consuelo.  Halo virial masses and virial radii are defined according to the spherical overdensity criterion of \cite{mvir_conv}; if a halo includes satellite halos within its virial radius, those are considered to contribute towards its virial mass.

\subsection{Kinetic and Potential Energy Calculation}

\label{s:grav_pot}

Calculating particle kinetic energies is straightforward ($\frac{1}{2}v^2$ per unit mass), where  $v$ is the physical velocity of the particle relative to the halo bulk velocity. The calculation of potential energies in expanding cosmologies poses normalization issues---the usual practice of setting the potential to zero at infinity is not well-defined.  The traditional method \cite{Rockstar,Klypin99,Springel01,Knollmann09,ASOHF10} is to ignore these issues and compute particle potentials in a Newtonian metric.  In this case, the potential $\phi$ of a point mass $M$ (i.e., the Green's function) is
\begin{equation}
\label{e:pot}
\phi(r) = -\frac{GM}{r}
\end{equation}
We discuss appropriate corrections for an expanding cosmology in later sections, but these are straightforwardly related to the Newtonian potential.  
%For completeness, we consider both this traditional definition as well as alternate definitions in \S \ref{s:fractions}, where $\phi=0$ for a chosen finite radius $R_0$; this latter case corresponds to substituting the following Green's function:
%\begin{equation}
%\label{e:pot2}
%\phi(r) = -\frac{GM}{r} + \frac{GM}{R_0}.
%\end{equation}
For calculating Newtonian gravitational potentials, we use a modified Barnes-Hut algorithm \citep{BarnesHut86}, as detailed in \cite{Rockstar}.

%In our analysis, we consider a range of particle energies, so it is helpful to define the ``boundedness'' $p$, of a particle as
%\begin{equation}
%\label{e:boundedness}
%b = \frac{KE-|PE|}{|PE|}
%\end{equation}
%where $KE$ and $PE$ are the kinetic and potential energies, respectively, of the particle.  A value of $p=-1$ indicates that the particle has no kinetic energy; a value of $p=0$ indicates that the particle has zero total energy (kinetic energy equals negative potential energy).  Values of $p>0$ indicate that the particle has positive energy.

%From the virial theorem, one might expect that the average value of $p$ within the virial radius of a halo should be $p=-0.5$.  However, this is not the case for several reasons: the force law in an expanding cosmology is not a simple $r^{-2}$ law (see Eq.\ \eqref{e:a}), not all particle orbits stay within the virial radius (see \S \ref{s:origins}), and the ``virial'' radius used here is set via a fixed overdensity threshold but cosmological halos are consistently perturbed away from virialization by mergers.  The combination of these factors makes the average value of $p$ within the virial radius of the halos in this study about -0.7 to -0.6.

\section{Unbound Particles in an Expanding Universe}

In this section, we first derive basic energy thresholds for unbound particles in an expanding universe and discuss basic qualitative features of their behavior (\S\ref{s:analytic}; see also Appendices \ref{s:boundedness} and \ref{s:num_sim} for more complete approaches).  We next present our operational definition of unbound particles in cosmological simulations in \S\ref{s:definition}.  Finally, we test how well kinetic energies can predict whether particles will become unbound in cosmological simulations in \S \ref{s:full_sims}.

\label{s:energy}

\subsection{Energy Thresholds and Qualitative Features}

\label{s:analytic}

\begin{figure}
\begin{center}
\vspace{-5ex}
\plotgrace{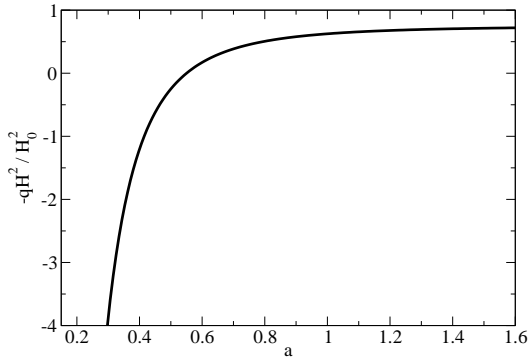}
\vspace{-5ex}
\end{center}
\caption{The evolution of the cosmological acceleration term $-qH^2$ with scale factor, relative to the Hubble parameter today ($H_0^2$).  Negative values imply that the Hubble flow decelerates escaping particles; escape to infinity is impossible while this is the case (see Eq.\ \eqref{e:a}).  Positive values of $-qH^2$ imply that the Hubble flow accelerates escaping particles.  The transition between these states occurs at $a=0.55$ for a flat cosmology with $\Omega_M = 0.25$ and $H_0 = 70$ km s$^{-1}$ Mpc$^{-1}$. In the far future, $-qH^2$ asymptotes to $\Omega_\Lambda H_0^2$.}
\label{f:qH2}
\end{figure}

If one considers a point mass in an expanding cosmology, the effective acceleration $\vec{a}_t$ experienced by a nonrelativistic test particle will be \citep{Nandra11}:
\begin{equation}
\label{e:a}
\vec{a}_t = \left(-\frac{GM}{r^2} -qH^2r\right)\hat{r}
\end{equation}
where $M$ is the value of the point mass, $H$ is the Hubble expansion rate ($\dot{a}a^{-1}$), and $q$ is the deceleration parameter ($-\dot{H}H^{-2} - 1 = -\ddot{a}a^{-1}H^{-2}$).  Both $q$ and $H$ evolve with time; the evolution of the product $-qH^2$ for our chosen cosmology is shown in figure \ref{f:qH2}.

At early times ($a<0.55$), $-qH^2$ is negative, and it is impossible for the test particle to escape to infinity.  This is because the deceleration rate increases linearly with distance: the farther the test particle gets, the stronger the deceleration it experiences.  However, at later times ($a>0.55$), $-qH^2$ becomes positive, and it becomes easier and easier for the test particle to escape.

The time-evolution of the force term in Eq. \eqref{e:a} means that energy is not conserved until the far future, when $-qH^2$ asymptotes to $\Omega_\Lambda H_0^2$ (see figure \ref{f:qH2} and \cite{Loeb11}).  The amount of kinetic energy loss (or gain) of the test particle until that time depends on its orbit.  Because a test particle launched at $a<0.55$ can orbit the point mass several times before escaping at $a>0.55$, \textit{there is not always a single energy threshold} for escape: certain orbit trajectories will have better resonance with the Hubble force term and experience larger kinetic energy changes than others.  However, for every location and initial direction of travel for the test particle, there is always a maximum kinetic energy threshold above which the particle is guaranteed to be unbound.

At very late times, the expansion of the universe becomes exponential (constant $H$), implying $q\to-1$.  Therefore, the product $-qH^2$ approaches a final value of $\Omega_\Lambda H_0^2$ (see also \cite{Loeb11}), as may be seen from the evolution of $H$ in a flat universe ($H^2(a) = H_0^2(\Omega_M a^{-3} + \Omega_\Lambda)$).  In this late-time case, it is possible to solve for the motion of the test particle analytically.\footnote{As may be seen in figure \ref{f:qH2}, this is a good approximation even at the present day; $qH^2=-0.63H_0^2$ at $z=0$ and changes only by $\sim 10\%$ in the next 3 Gyr.}  There is an equivalence radius where the ``push'' of the Hubble flow balances the pull of gravity:\footnote{If $M$ corresponds to the mass of a dark matter halo, $r_e = R_h \sqrt[3]{\frac{\Delta_h}{-2q}}$, where $R_h$ is the halo radius, and $\Delta_h$ is the overdensity which is used to define $R_h$.  Thus, the equivalence radius is typically 3-5 times the virial radius for $a\ge1$.}
\begin{equation}
\label{e:r_e}
r_e = \sqrt[3]{\frac{GM}{-qH^2}},
\end{equation}
If the particle can travel beyond this equivalence radius, it will escape; otherwise, it will remain bound.  The minimal velocity threshold for escape on a radial path is easy to find:\footnote{Due to Gauss's law, this formula works for any spherical matter distribution contained within $r_0$.}
\begin{equation}
\label{e:b_limit}
v_0^2 > \frac{2GM}{r_0} + qH^2 \left[r_0^2 - 3 r_e^2\right]
\end{equation}
where $r_0$ and $v_0$ are the test particle's initial radius and speed.  When $-qH^2 > 0$ at late times, the velocity required to escape is less than in the Newtonian case ($\sqrt{\frac{2GM}{r_0}}$).  Deriving the escape threshold for non-radial paths is substantially messier as it involves a quartic equation (see Appendix \ref{s:boundedness}).  However, it is a fact that \textit{escape along a tangential trajectory requires more energy than escape along a radial trajectory}.  Nonzero initial angular momentum (i.e., at least a partially tangential trajectory) means that the angular momentum at the equivalence radius will also be nonzero.  Thus, it is impossible to convert all the initial kinetic energy into potential energy as in the radial case, requiring slightly more energy to cross the escape threshold radius $r_e$.

\subsection{Defining ``Unbound'' Particles in Cosmological Simulations}

\label{s:definition}

The ability to trace particles into the far future provides a simple way to classify which particles are ``bound'' or ``unbound.''  At late times, the exponential expansion of the universe means that halos become widely separated \cite{Busha03}.  They also become more spherical, meaning that the gravity---Hubble flow equivalence radius $r_e$ (Eq.\ \eqref{e:r_e}) is an excellent approximation to halo boundaries.\footnote{As may be readily verified, using $r_e$ as the halo radius at late times is identical to defining halos using a spherical overdensity criterion of $2\Omega_\Lambda$ times the critical density.}

At a given redshift $z$, every halo in the simulation we use can be associated with a descendant halo at late times ($a=100$ in our case), defined as the halo which shares the largest number of particles.  Then, particles in a halo at redshift $z$ can then be checked against the descendant halo.  If a particle is within $r_e$ of the descendant at $a=100$, it is considered bound; otherwise, it is considered unbound.

The main remaining source of ambiguity is halo mergers: if a smaller halo will eventually merge into a larger one, it is unclear how to classify its particles.  Many high-energy particles from the smaller halo will be captured by the larger halo even though, in isolation, they would have readily escaped from the smaller halo.  It is somewhat unappealing to call these high-energy particles ``bound,'' even though they will never escape to infinity.  An instantaneous measure of boundedness, such as total kinetic and potential energy, would provide a way around this issue; however, as we discuss in the next section, the use of total energies introduces more problems than it solves.  For this reason, we provide in later sections separate estimates of unbound particle fractions for halos which are the most-massive progenitors of their descendants at $a=100$.

\subsection{Kinetic and Potential Energies as an Unbound Classification Technique}
\label{s:full_sims}

\begin{figure}
\begin{center}
\vspace{-5ex}
\plotgrace{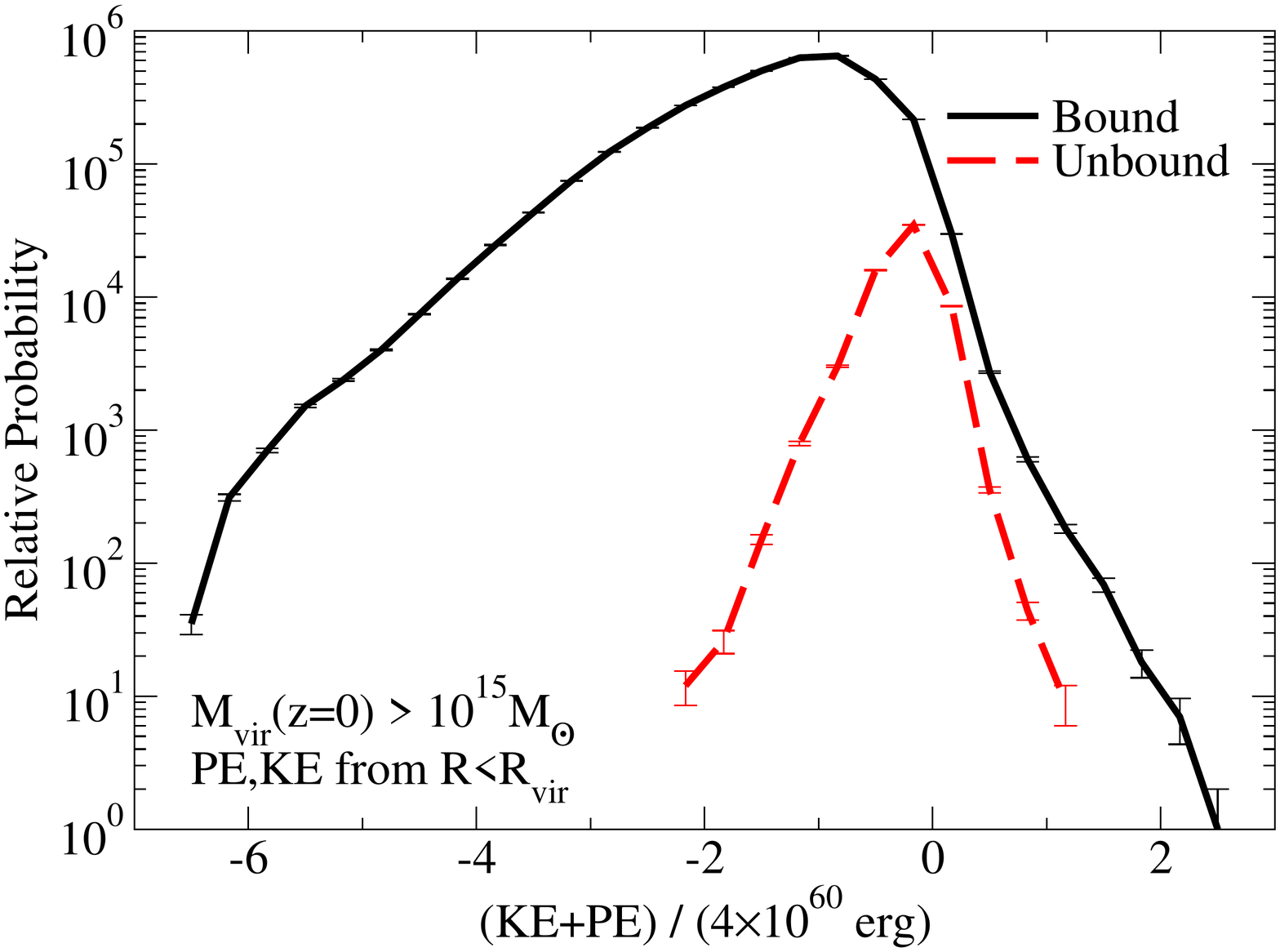}\plotgrace{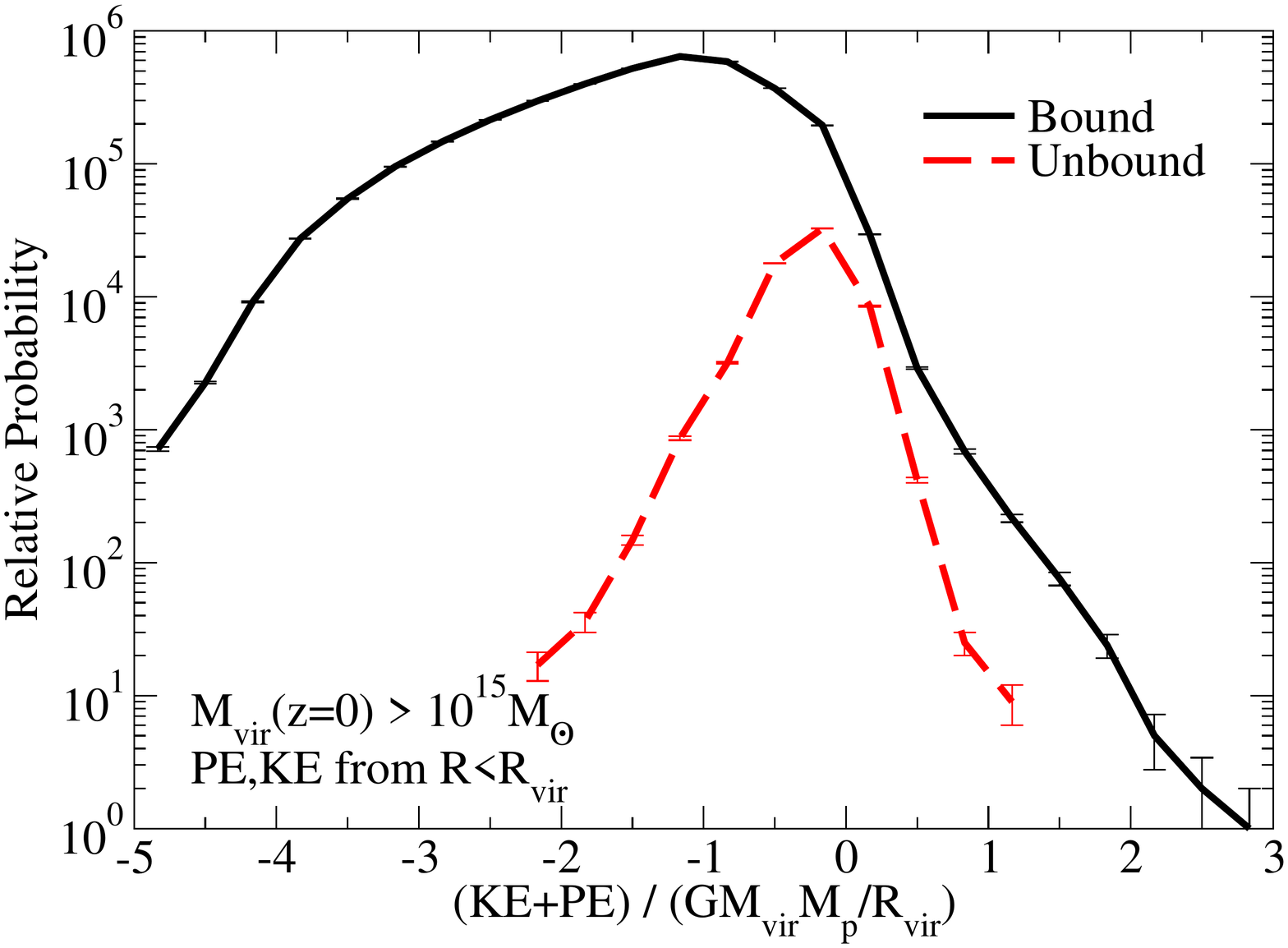}\\[-3ex]
\plotgrace{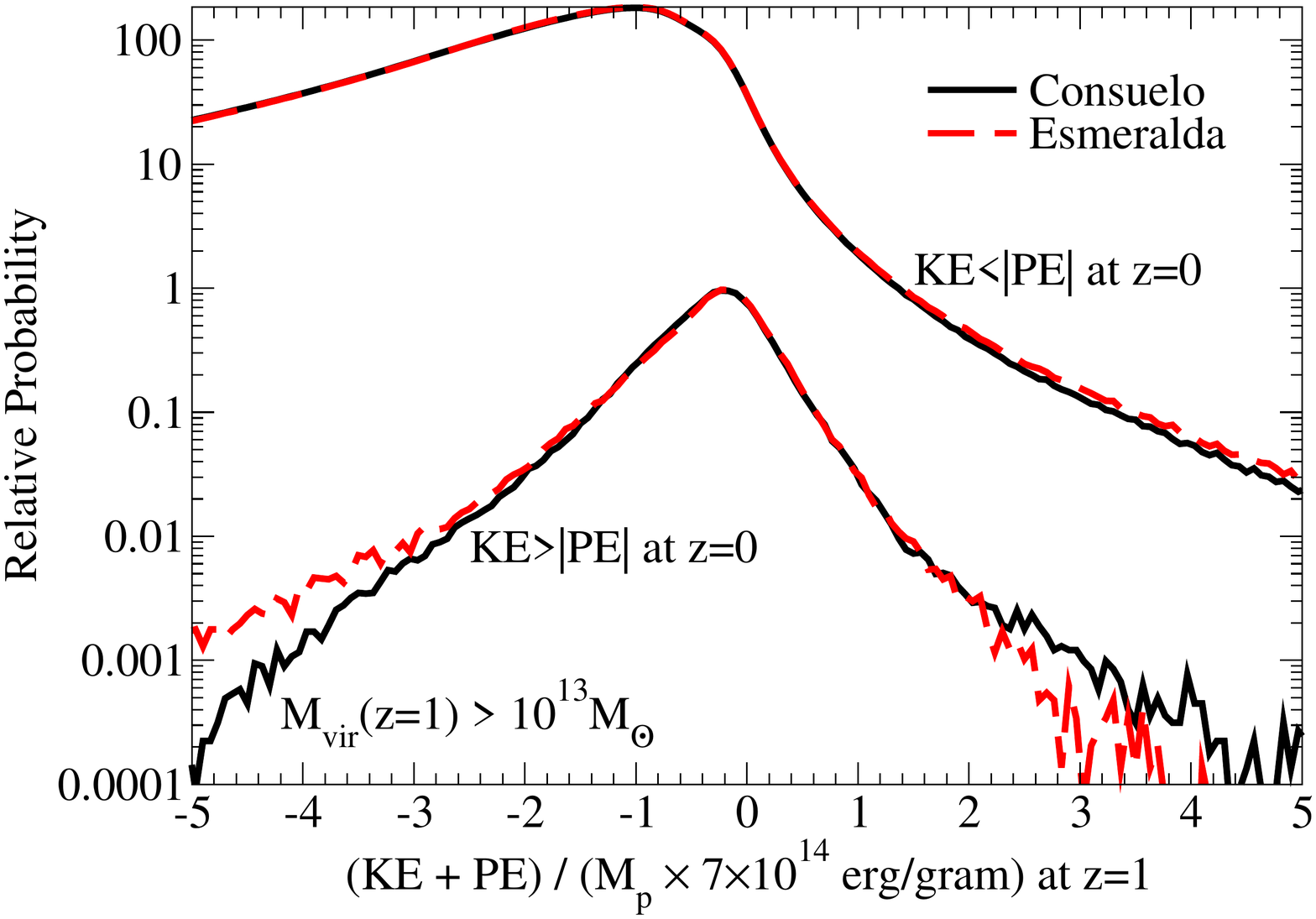}\plotgrace{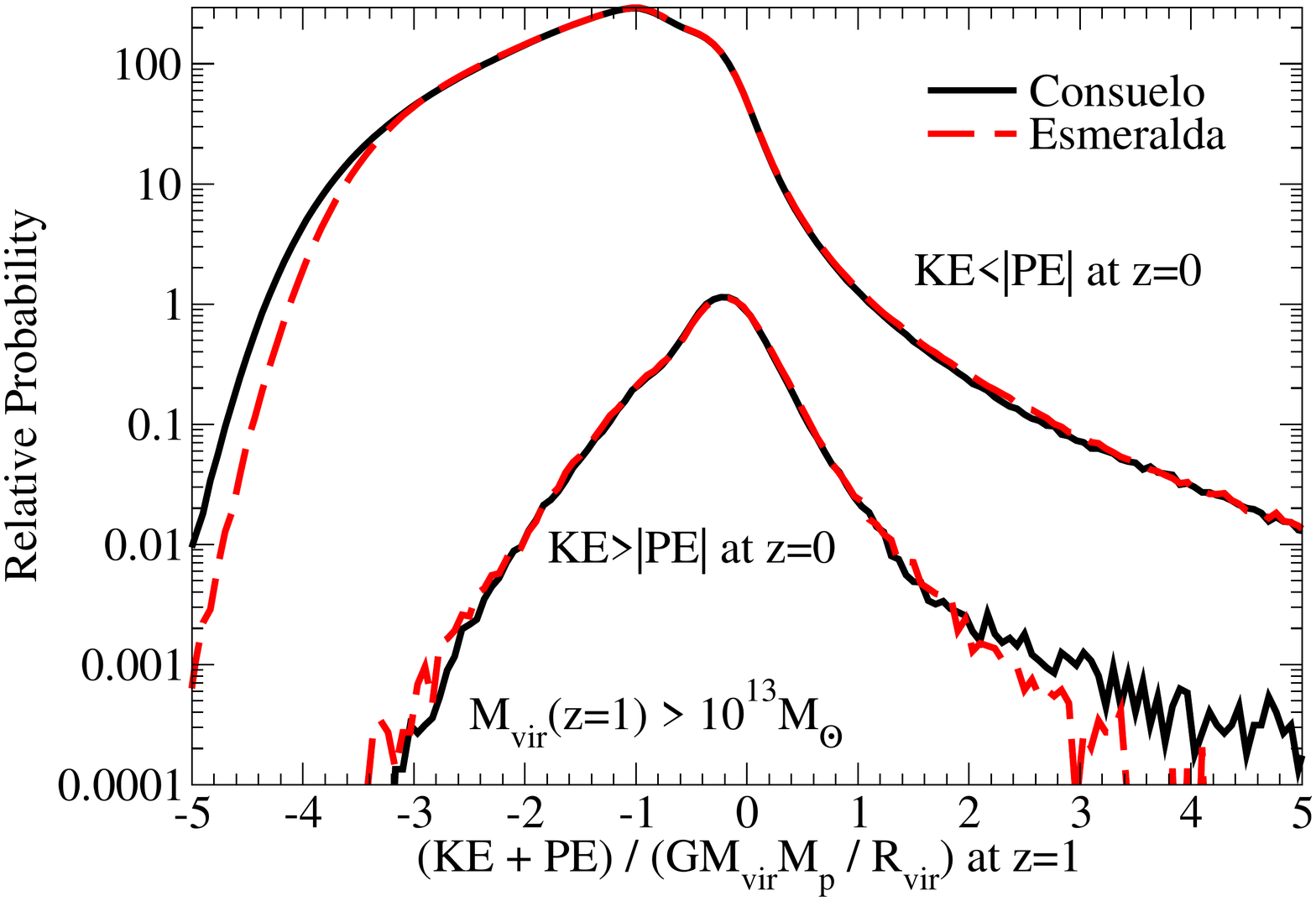}
\vspace{-5ex}
\end{center}
\caption{\textbf{Top-left panel}: the kinetic energy plus potential energy ($KE+PE$) distribution at $z=0$ of all particles in $\Mvir > 10^{15}\Msun$ halos which become unbound by $a=100$.  No energy threshold is capable of reliably distinguishing bound from unbound particles.  \textbf{Top-right panel}: same, except normalized by the potential energy of a particle at the halo virial radius.  \textbf{Bottom panels}: the $KE+PE$ distribution at $z=1$ for particles with $KE>|PE|$ and particles with $KE<|PE|$ at $z=0$.  As above, the left panel shows the direct $KE+PE$, and the right panel shows $KE+PE$ normalized to the potential energy of a particle at the virial radius at $z=1$.  $M_p$ represents the particle mass, which is $10^{11}\Msun$ in the simulation in the top panel, $1.33\times 10^{10}\Msun$ for the \textit{Esmeralda} simulation, and $2.7\times 10^{9}\Msun$ for the \textit{Consuelo} simulation.  See text for details.  }
\label{f:energy_thresh}
\end{figure}

Particle motion in cosmological halos is more complex than for the point-mass models described in \S \ref{s:analytic}.  Some additional effects include particles scattering off of substructure, tidal disruption of substructure, halo mass accretion, nonspherical halo mass profiles, and halo mergers.  These phenomena all result in dark matter particles exchanging energy with the environment in nontrivial ways.  Therefore, because particle kinetic energies have been used almost exclusively in the literature (our previous papers included) to classify bound and unbound structure \cite{Rockstar,Knollmann09,ASOHF10,Klypin99,Genel09,Wang11}, it is important to test the validity of this technique.

Since we can directly classify boundedness by tracing particle trajectories, it is straightforward to plot the energy distribution of unbound particles at $z=0$.  We do so for $\Mvir>10^{15}\Msun$ halos in our main simulation (which are resolved with $>$10,000 particles) at $z=0$.  While these halos are almost always the largest collapsed structures in their nearby environment, we exclude the few which merge into even larger halos from this test.

As the top panels of figure \ref{f:energy_thresh} show, we find that \textit{no} energy threshold at $z=0$ is able to accurately predict which particles will have escaped the halo by $a=100$.  \textit{No matter} what a particle's kinetic energy is at $z=0$, it is more likely to be bound than unbound.  We have also tested increasing the radius for particles included in the potential energy calculation to $2\Rvir$ and $3\Rvir$; the identical conclusion applies.
 
Because we only have a single simulation run to $a=100$, it is very difficult to perform a direct resolution test of this conclusion.  Instead, we perform a simpler test, which is to check how well particle total energies ($PE+KE$) at $z=1$ correspond to those at $z=0$ for both the \textit{Esmeralda} and \textit{Consuelo} simulations.  In both simulations, we categorize particles depending on whether they have $KE>|PE|$ or $KE<|PE|$ at $z=0$.  Then, we plot the distributions of $KE+PE$ for both sets of particles at $z=1$.  To avoid small particle-number effects, we exclude halos below $10^{13}\Msun$ (780 and 3700 particles for Esmeralda and Consuelo, respectively).  As before, we only consider halos at $z=1$ which are the most-massive progenitors of halos at $z=0$, so as to exclude cases where a smaller halo merges into a larger one.  Finally, we exclude halos which lose mass between $z=1$ and $z=0$, as these are most likely experiencing tidal forces from a nearby larger halo which could influence particle trajectories.

As shown in the bottom panels of figure \ref{f:energy_thresh}, particle total energies are very much not conserved between $z=1$ and $z=0$.  This makes the previous conclusion even stronger: particle energies at $a=0.5$ are in general a poor predictor of particle energies at $a=1$, let alone at $a=100$.  That said, \textit{highly-bound} particles at $z=1$ and $z=0$ do seem much more likely to remain bound at later times, meaning that highly-bound particles such as stars are unlikely to escape entirely from halos in mergers (see also \S \ref{s:hostless} and \S \ref{s:baryons}). 

Having done many tests over shorter time intervals (not shown), we remark that the decoherence of kinetic energies over time is a generic feature of dark matter particles.  However, this is not necessarily the case for high-velocity objects such as hypervelocity stars.  We derive how the standard kinetic energy threshold for escape changes when properly accounting for the Hubble expansion as well as halo mass accretion in Appendix \ref{s:num_sim}.  In these cases, when the kinetic energy source is not large-scale gravitational forces, then the effect of interactions with substructure is expected to be small (see Appendix \ref{s:substructure}).

\begin{figure}[tp]
\plotminigrace{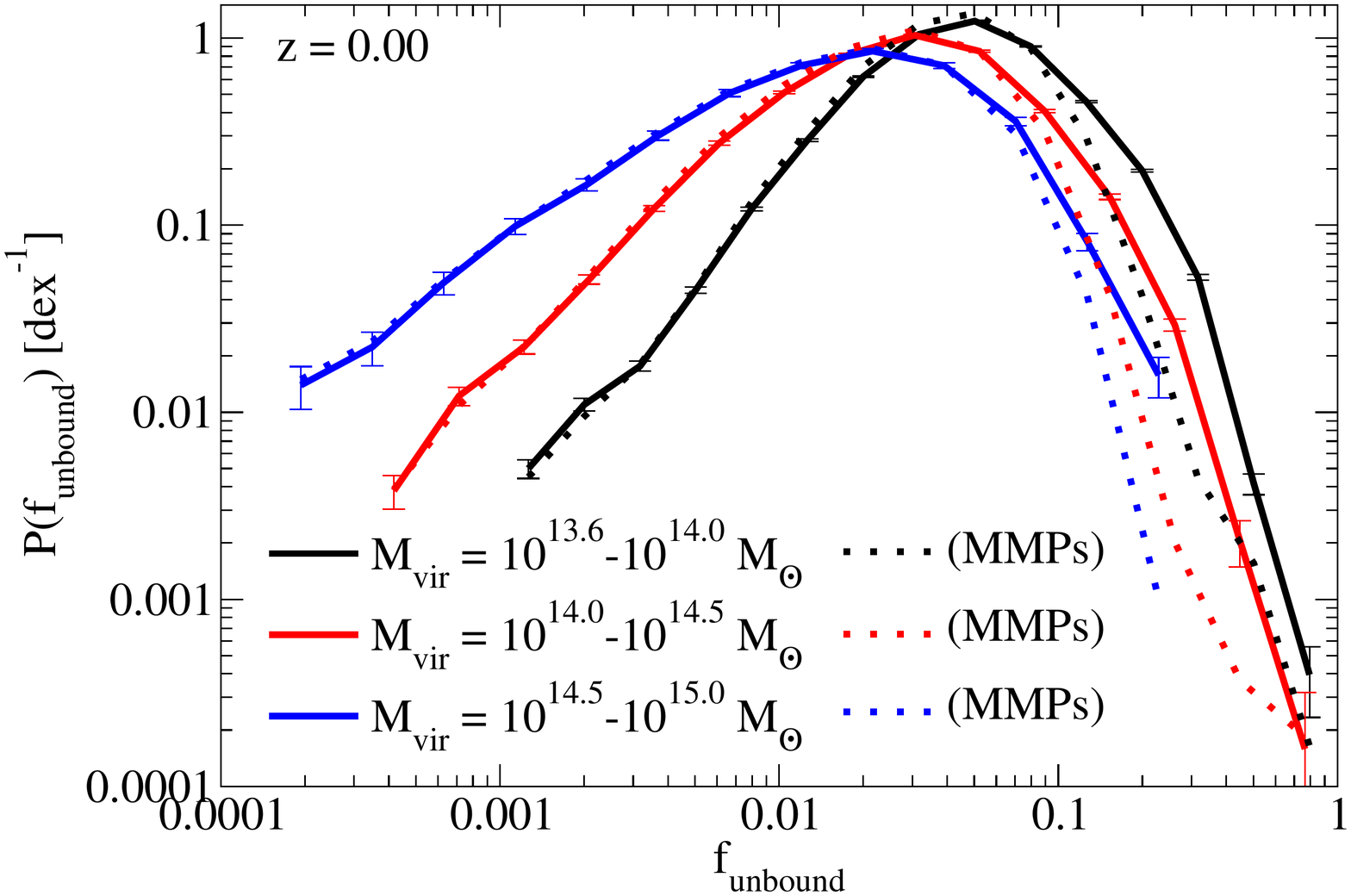}
\plotminigrace{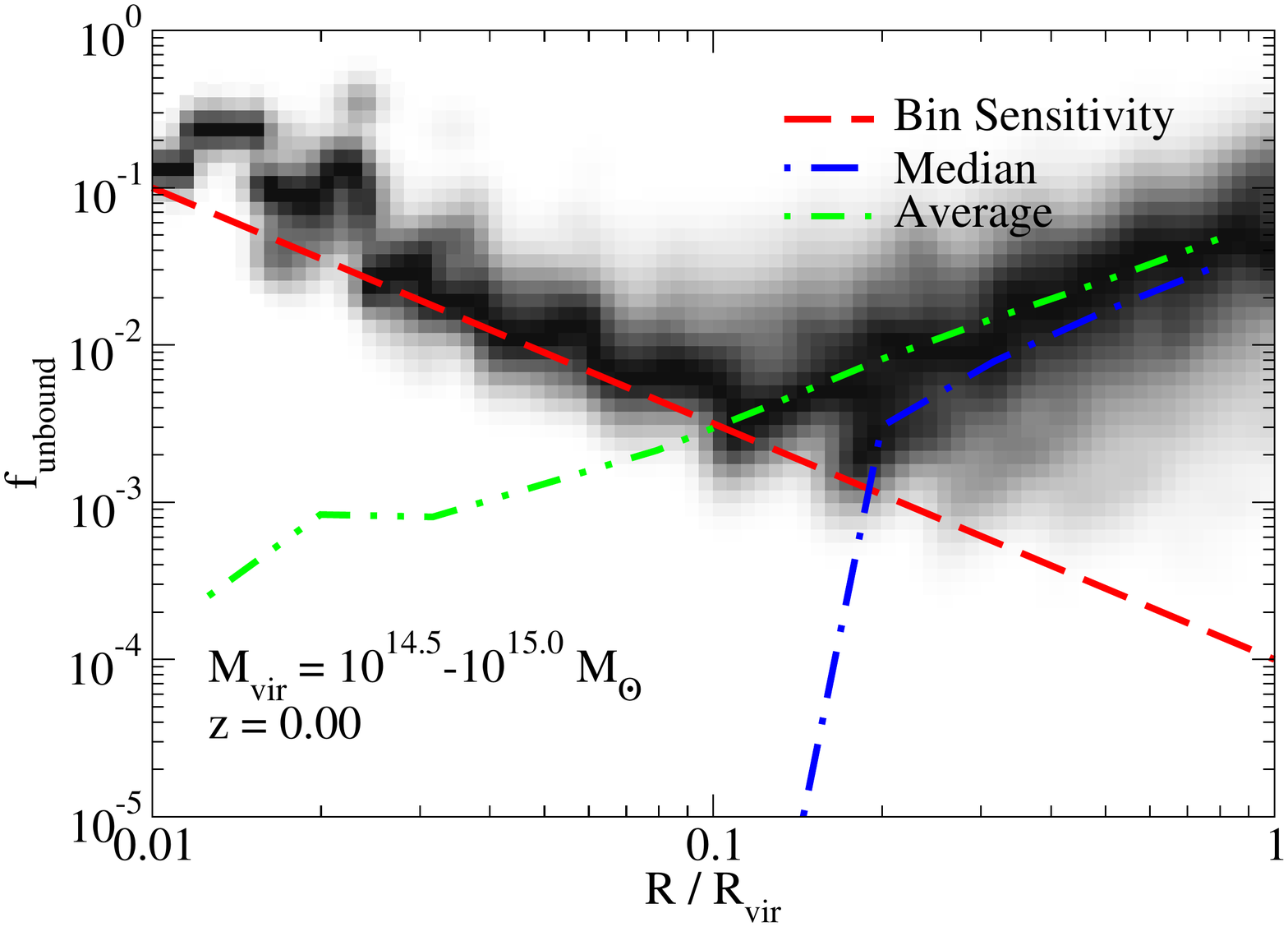}
\caption{The \textbf{left panel} shows the probability distribution of unbound mass fractions in halos at $z=0$, derived from tracing particle trajectories to $a=100$.  This averages 2-6\% for the halos in question, with significant mass dependence.  The \textit{dotted} lines show the probability distribution for the halos at $z=0$ which are the most-massive progenitors (MMPs) of halos at $a=100$.  The \textbf{right panel} shows the conditional density distribution of the unbound particle fraction as a function of radius in massive ($10^{14.5}\Msun < \mvir < 10^{15}\Msun$) halos at $z=0$.  The \textit{red dashed line} shows the limit of one particle per radial bin; effectively, the individual halo sensitivity limit.  Because the unbound fraction as a whole passes below this limit at small radii, the median unbound fraction (\textit{blue dot-dashed line}) rapidly drops to zero from the floor effect.  On the other hand, the average unbound fraction (\textit{green double dot-dashed line}) follows the trend established at higher radii down to a small fraction of the virial radius.
}
\label{f:p_dist}
\end{figure}

\begin{figure}[tp]
\vspace{-15ex}
\plotgrace{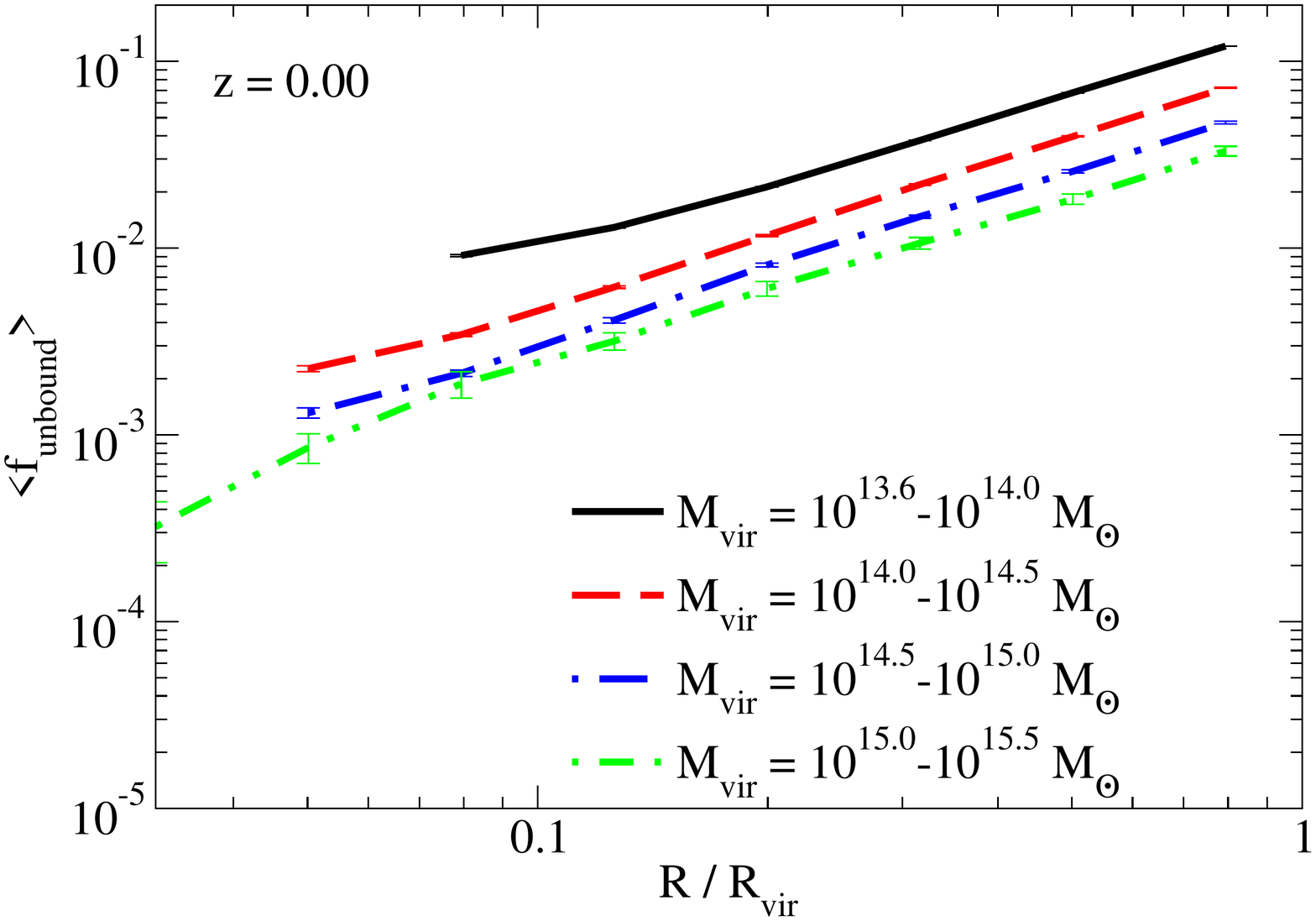}\plotgrace{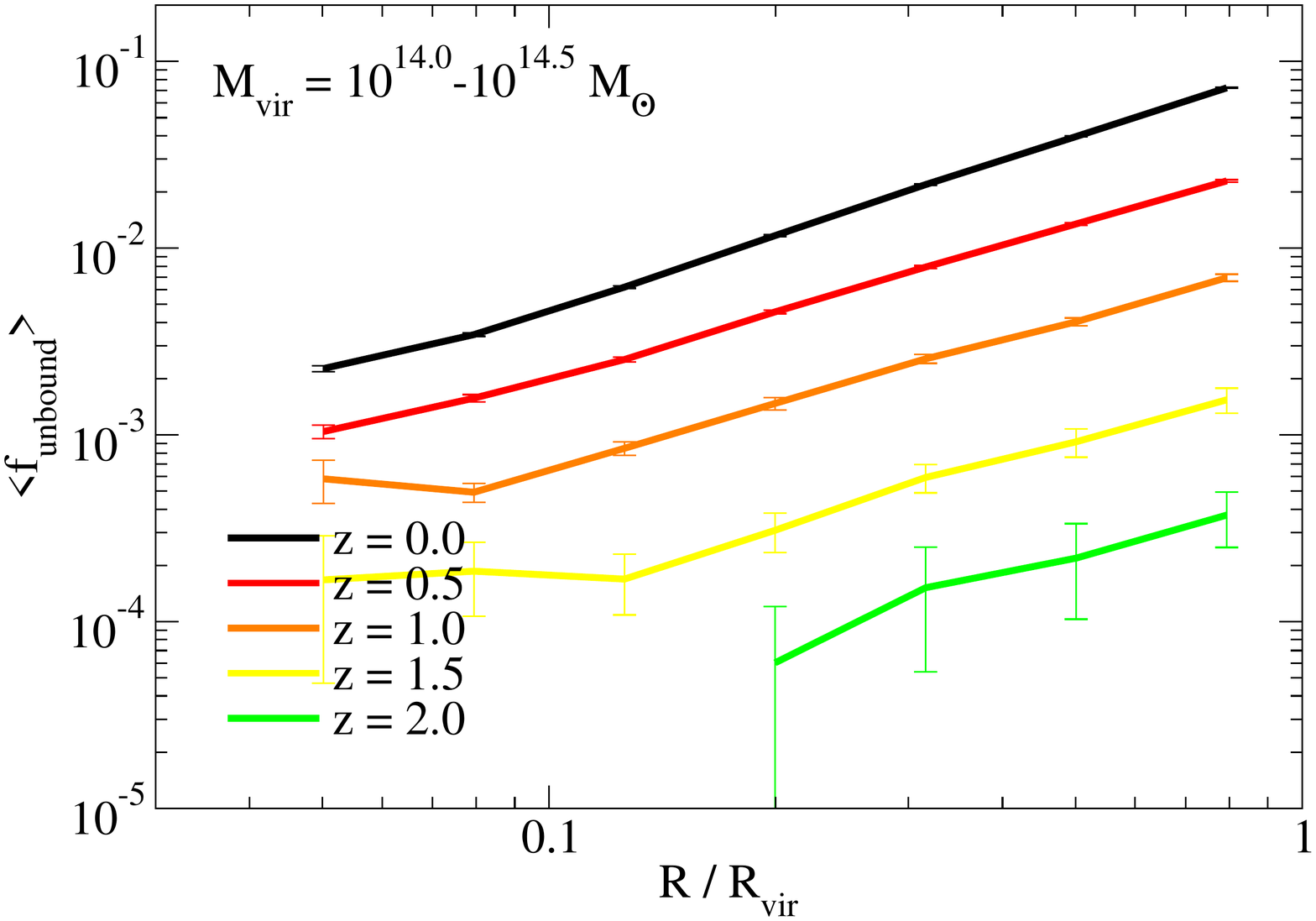}\\[-4ex]
\plotgrace{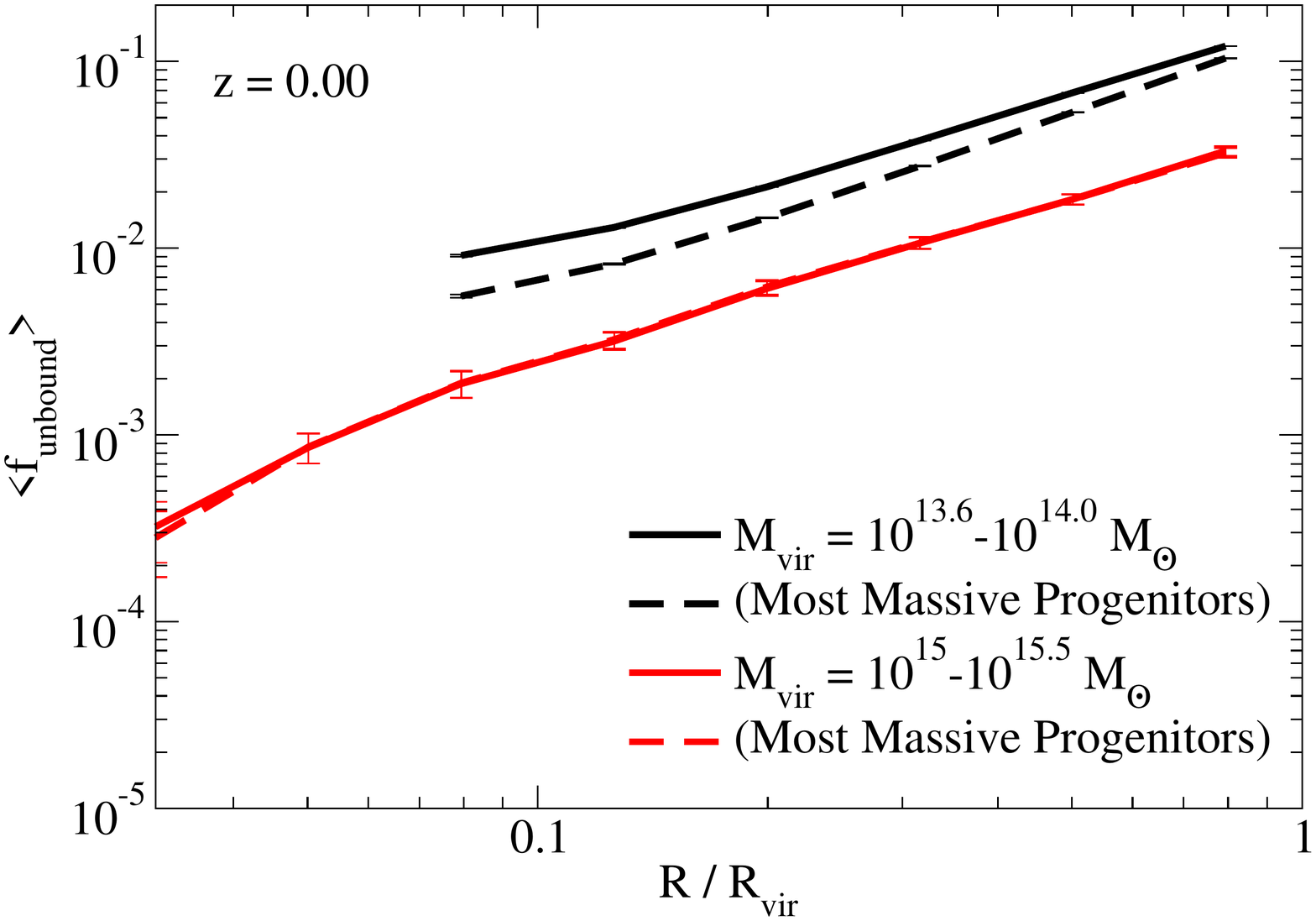}\plotgrace{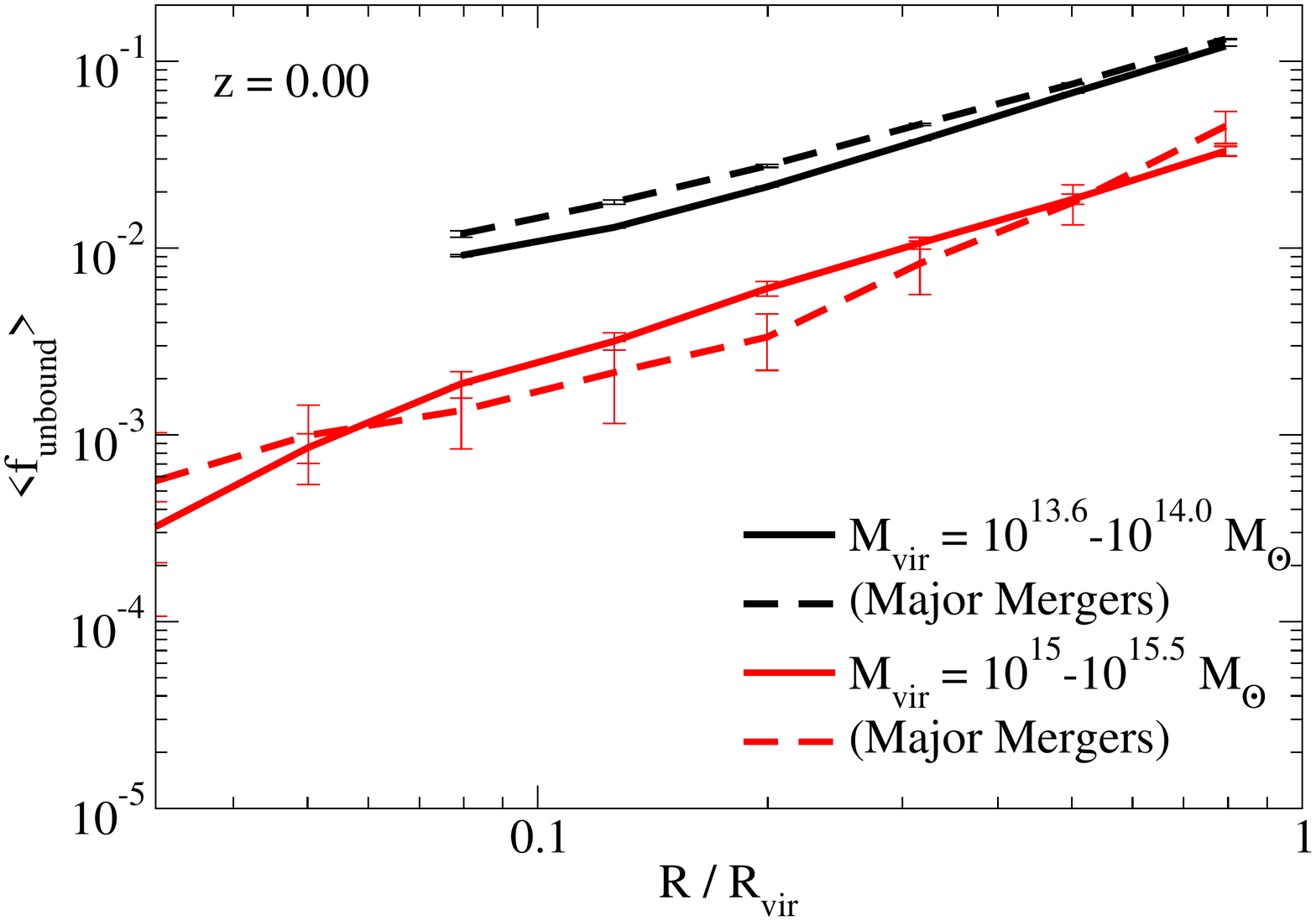}\\[-4ex]
\caption{\textbf{Top-left panel}: The average unbound fraction as a function of mass and radius; smaller halos have more unbound particles.  For this and other panels, limited resolution prevents calculating unbound fractions within 50 h$^{-1}$ kpc of halo centers.  \textbf{Top-right panel}: The unbound fraction as a function of radius and redshift for $10^{14}-10^{14.5}\Msun$ halos.  The unbound fraction drops off precipitously at higher redshifts.  \textbf{Bottom-left panel}: The unbound fraction for all halos as compared to halos which are the most-massive progenitors of halos at $a=100$.  Most-massive progenitors have fewer unbound particles than average; at higher masses, more halos are most-massive progenitors, so this effect is less apparent. \textbf{Bottom-right panel}: The unbound fraction for all halos as compared to halos undergoing major mergers.  Halos in major mergers have surprisingly few extra unbound particles, on average.  Error bars for all panels are calculated using jackknife statistics.}
\label{f:radial}
\end{figure}

\section{Properties of Unbound Particles}

\label{s:results}

As discussed in the previous section, total kinetic and potential energy is a poor classifier of unbound dark matter particles.  Instead, we turn to a simulation run through the far future ($a=100$) and trace particle and halo trajectories.  We present statistics of the fraction of unbound particles in halos in \S \ref{s:radial}, as well as the variation with respect to radius, redshift, and halo mass.  We also discuss variations in the unbound fraction with environment in \S \ref{s:env_dep}.

\subsection{Fractional Occurrence of Unbound Dark Matter Particles in Halos}

\label{s:radial}

Individual host halos can have a wide variety of unbound particle fractions at $z=0$, as shown in the left panel of figure \ref{f:p_dist}.  On average, about 2-6\% of particles will escape to infinity over the mass range considered ($10^{13.6}\Msun < \mvir < 10^{15} \Msun$), with significant mass dependence.  There is also a strong dependence on whether a halo is a most-massive progenitor (MMP) of the descendant halo at $a=100$ or not.  On average, a non-MMP has an unbound fraction 2-3x larger than an MMP halo of similar mass.  Indeed, we note that some halos in the top panels of Fig.\ \ref{f:p_dist} can have very high unbound fractions, approaching 100\%.  These halos end up being shredded into many pieces at $a\gg 1$.  Two out of the $\sim$76,000 halos in the mass range $10^{13.6}\Msun < \mvir < 10^{14} \Msun$ ended up being shredded into more than 5 pieces, each no larger than 20\% of the original mass.

The right panel in figure \ref{f:p_dist} shows the spread in unbound fractions as a function of radius for massive halos ($10^{14}\Msun < \mvir < 10^{14.5} \Msun$) at $z=0$.  The significant scatter seen in figure \ref{f:p_dist} remains at all radii.  However, at small radii ($<0.2\rvir$), the average unbound fraction is so low that most halos will have less than a single unbound particle per radial bin.  As a consequence, the median unbound fraction across halos significantly underestimates the overall trend at small radii.  For that reason, we express radial unbound fractions in terms of the average across all halos, which better follows the trend of unbound fractions established at larger radii.

The average radial profile of the unbound fraction (i.e., the ratio of unbound to total particles in radial bins) is shown in figure \ref{f:radial}.  In all cases, the unbound fraction increases roughly linearly with radius.  For an NFW \cite{NFW97} halo density profile, this would imply that the physical density of unbound particles peaks at the halo center.

The top-left panel in figure \ref{f:radial} shows how the radial dependence of the average unbound fraction depends on halo mass.  As in the top panel of figure \ref{f:p_dist}, smaller halos have larger unbound fractions across all radii.  For smaller halos in this panel, the radial dependence is truncated at 50 h$^{-1}$ kpc, as the gravitational accuracy of the simulation degrades rapidly below that radius.

The top-right in figure \ref{f:radial} shows the redshift dependence of the radial unbound fraction, which has substantial evolution from $z=0$ to higher redshifts.  This trend is preserved across all the mass ranges we consider, and matches the qualitative expectation that the deceleration of the universe at early times makes it much harder for particles to escape (see Appendix \ref{s:num_sim}).

The bottom-left panel in figure \ref{f:radial} shows how the unbound fraction changes if only most-massive progenitor halos are selected.  As expected from figure \ref{f:p_dist}, MMP halos have lower unbound fractions at all radii, as compared to average halos of the same mass.  However, this difference is less pronounced for more massive halos, as the overall fraction of halos which are MMPs increases with halo mass.

The bottom-right panel in figure \ref{f:radial} shows how the unbound fraction changes if only halos undergoing major mergers are selected.  Specifically, a halo is considered to be undergoing a major merger if it contains a satellite halo with $v_\mathrm{max,sat} > 0.6v_\mathrm{max,host}$, where $v_\mathrm{max}$ is the maximum circular velocity, corresponding approximately to an infalling mass ratio of 0.3:1.  This panel shows that, for massive halos, the presence of a major merger has very little impact on the unbound fraction.  However, for less-massive halos (e.g., $10^{13.5}\Msun < \mvir < 10^{14.0}\Msun$), a major merger can increase the unbound fraction somewhat more (by 5--20\% overall).  At first glance, it may appear surprising that halos undergoing major mergers do not have substantially more unbound particles.  However, the mass in an incoming major merger gets added to that of the main host halo.  Hence, while the incoming particles may have been unbound relative to the potential of the original merging halo, they are mostly bound relative to the \textit{combined} gravitational potential of the incoming and original particles within the virial radius.  This effect is explicitly shown in \S \ref{s:env_dep}.

To summarize the most important trends, the unbound fraction is a strong function of radius.  The unbound fraction on galaxy scales (within $0.05 \rvir$) is below 1\%, whereas it can reach 10\% or more at the halo radius.  Major mergers, lower halo masses, and lower redshifts all correlate with increased unbound fractions, but these effects are comparatively subdominant.

\begin{figure}
\plotgrace{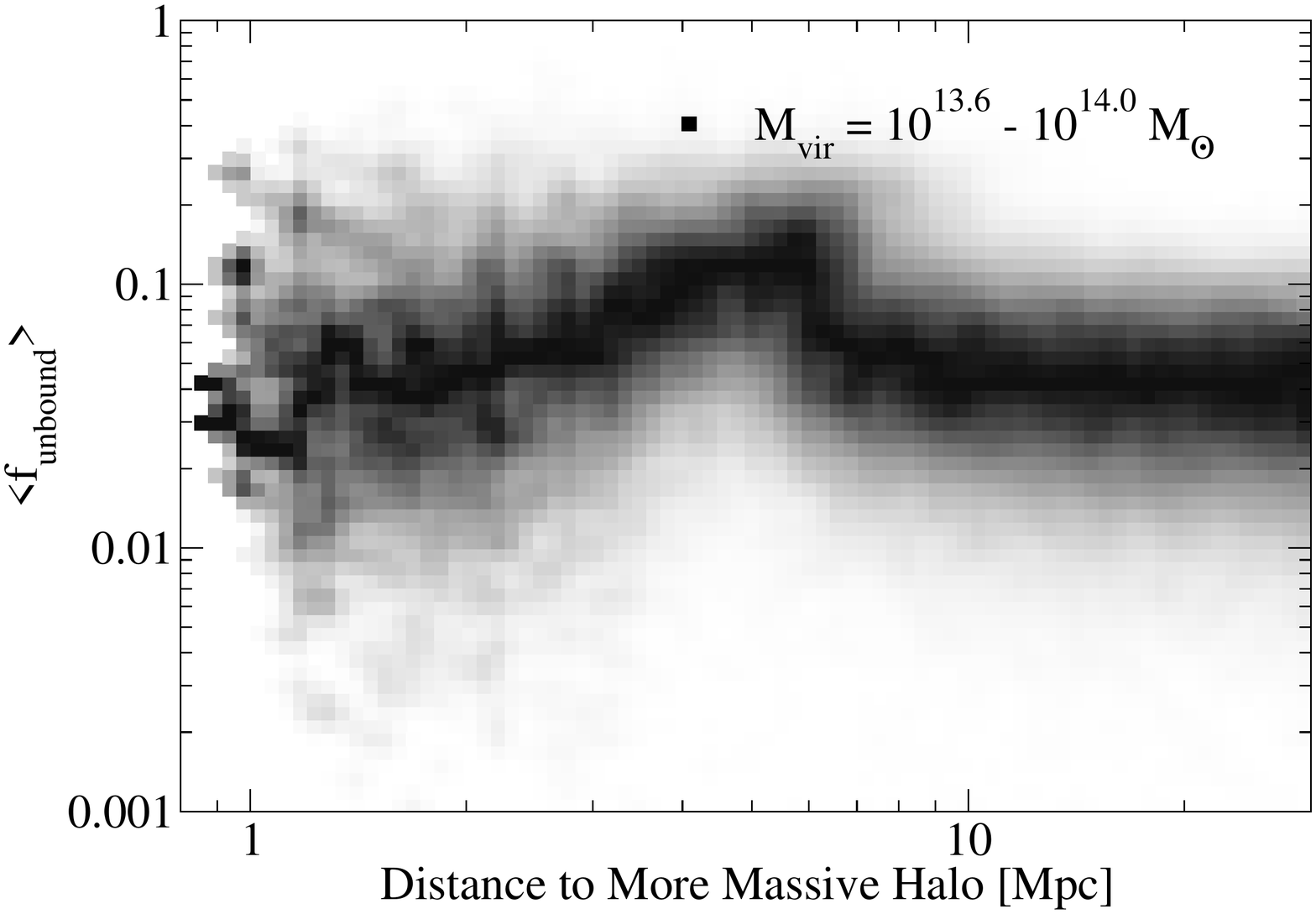}\plotgrace{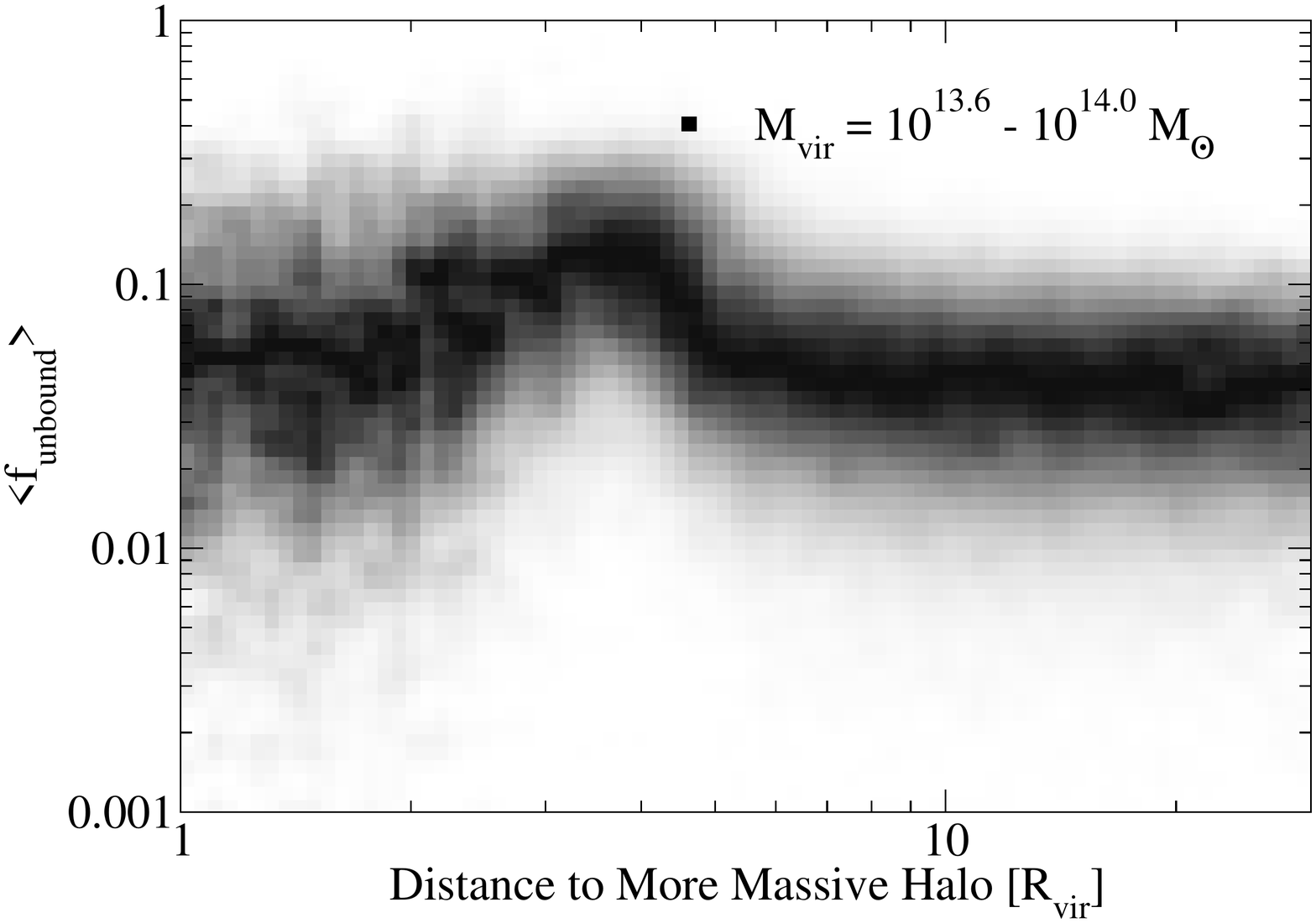}\\[-5ex]
\begin{centering}
\plotgrace{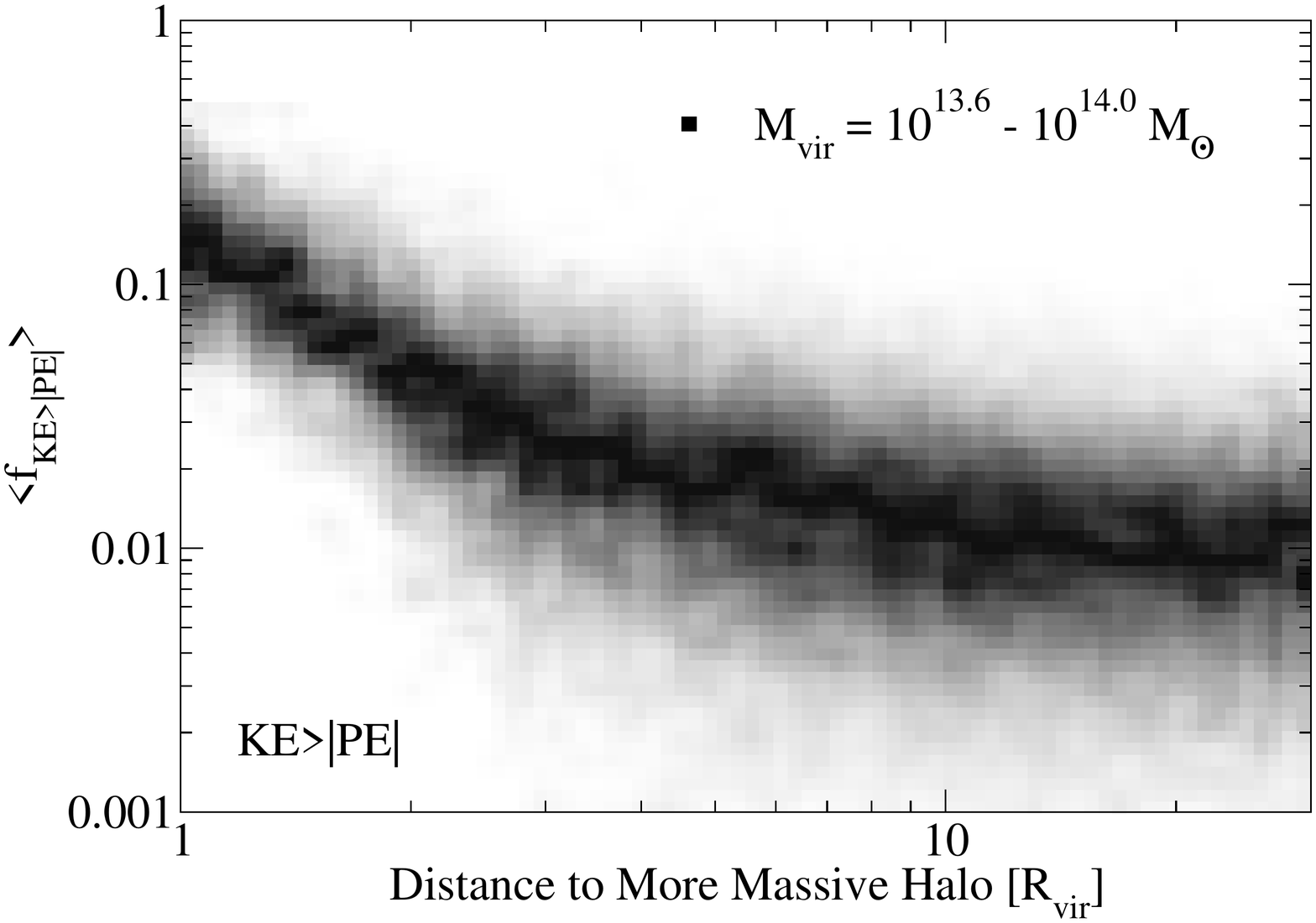}\\[-4ex]
\end{centering}
\caption{\textbf{Top-left panel}: The conditional probability distribution of the unbound fraction in $10^{13.6}-10^{14}\Msun$ halos as a function of the distance to the nearest larger halo at $z=0$.  \textbf{Top-right panel}: Same as top-left panel, except now in units of the larger halo's virial radius.  This shows that the unbound fraction for smaller halos climbs steeply within 3-4$\rvir$ of a larger halo.   \textbf{Bottom panel}: For comparison, the fraction of particles in halos with kinetic energies larger than the magnitude of their potential energies ($KE>|PE|$).}
\label{f:env}
\end{figure}

\subsection{Environmental Dependence}

\label{s:env_dep}

There are several ways of quantifying the local environment; one common method is to use the distance to the nearest larger halo.  As shown in the top-left panel of figure \ref{f:env}, the unbound fraction for $10^{13.6}-10^{14}\Msun$ halos has a clear maximum when a larger halo is 5 Mpc away.  In units of the larger halo's virial radius, this occurs between 3 and 4 $\Rvir$ (top-right panel of figure \ref{f:env}).  This maximum persists for more massive halos as well (not shown).

This feature has a straightforward explanation.  When larger halos are much more than $3-4\Rvir$ away, the Hubble flow is stronger than the influence of gravity (see Eq.\ \eqref{e:r_e}); this will accelerate away the larger halos before they have the chance to interact strongly.  When a larger halo is around $3-4\Rvir$ away, the smaller halo is just barely within the region which will eventually collapse onto the larger halo; hence, particles in the smaller halo can more easily escape.  Finally, when there is a larger halo very nearby to the smaller halo, then most of the particles in the smaller halo will remain bound to the larger halo's gravitational potential.

This last point reinforces the main issue with our adopted definition for unbound particles.  Smaller halos merging into larger ones will have most of their particles considered as ``bound'' even if, in isolation, they would have escaped from the smaller halo.   The bottom panel of figure \ref{f:env} shows this clearly---the fraction of high-energy ($KE>|PE|$) particles in halos increases dramatically when a larger halo is nearby.  That said, the source of many of these high-energy particles is the larger halo itself, meaning that their association with the smaller halo is just as much an issue with the spherical overdensity definition of a halo as it is with the definition of unbound particles.

\section{Discussion and Implications}

\label{s:discussion}

The most important radius for the individual trajectory of dark matter particles is the equivalence radius (Eq.\ \eqref{e:r_e}), beyond which the Hubble flow will accelerate the particle away from the gravitational pull of the halo.  At early times ($a<0.55$), the deceleration of the universe makes it impossible for particles to completely escape from halos.  However, at late times, the acceleration of the universe brings the equivalence radius from infinity down to roughly 4.5 times the virial radius.  This leads to the creation of ``islands'' at late times ($a\gg 1$), whereby particles (and, indeed, other halos) which are not within the equivalence radius are expanded away at rates that gravity cannot overcome (see also \cite{Busha05,Loeb11}).

Perhaps surprisingly, we have shown that total kinetic and potential energy is a poor predictor of whether particles will become unbound in cosmological simulations (\S \ref{s:full_sims}).  Instead, the only robust way of determining which particles are unbound is to test whether they leave the equivalence radius in the far future.  We have shown that the fraction of unbound particles defined this way  is a strong function of distance to the halo center as well as redshift (\S \ref{s:radial}).  The fraction of unbound particles also depends somewhat on the distance to the nearest larger halo, although the fraction of high kinetic-energy particles has a much stronger environmental dependence (\S \ref{s:env_dep}).

We note that the total kinetic and potential energy should not fare any better as a predictor for boundedness in satellite halos.  Over the long term, this is clear because most satellites will dissipate into their hosts---i.e., none of the particles will remain bound to each other.  Even over the short term, the total kinetic and potential energy ignores the tidal force from the host halo, which can be significant compared to the gravitational force from the satellite halo.  That said, removing particles with positive total kinetic and potential energy from satellite halos does represent a convenient cut in phase space for \textit{categorizing} particle membership.  Many other possible cuts exist \citep[e.g.,][]{Rockstar,Maciejewski09,Elahi11}.

In the remainder of this section, we discuss several implications of the previous results on unbound particles.  The dependence of particle kinetic energies on halo environment suggests a possible implication for direct dark matter detection experiments, which we test in \S \ref{s:dm_detect}.  The radial dependence of the unbound particle fraction is interesting in connection to the kinetic energy source for hypervelocity supernovae and GRBs, which we discuss in \S \ref{s:hostless}.  Finally, the possibility of escaping dark matter particles raising the baryon fraction in halos is discussed in \S \ref{s:baryons}.

\subsection{The Effect of Environment on Dark Matter Detection}

\label{s:dm_detect}

\begin{figure}
\begin{centering}
\plotgrace{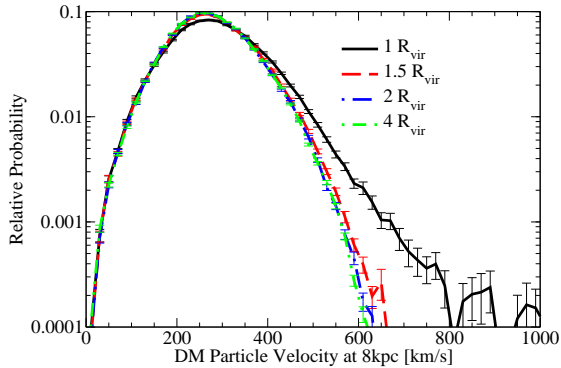}
\caption{The velocity distribution function (VDF) for Milky Way-sized halos at the Earth radius as a function of environment.  Lines in the plot show the VDF for halos binned by the distance to the nearest larger halo, in units of the larger halo's virial radius.  (This serves as a proxy for the tidal force).  Very little difference is seen except for halos currently merging with a larger halo.  Andromeda ($\sim3\Rvir$ away) therefore has minimal effect on the VDF near the Earth radius.}
\end{centering}
\label{f:vdf}
\end{figure}

For the Milky Way, the nearest larger halo would be associated with M31 (Andromeda), which is approximately three virial radii away \cite{Ribas05}.  Assuming an NFW profile, the approximate one-sigma mass range of the Milky Way is $5.4\times 10^{11}\Msun < \mvir < 1.4 \times 10^{12}\Msun$ \cite{Smith07,Busha11,Xue08}, the galactocentric distance is $\sim$8 kpc \cite{Honma12}, and the circular velocity at the Earth's radius is $\sim$238 km/s \cite{Honma12}.  Selecting halos in the one-sigma mass range from the Bolshoi simulation, we calculate the velocity distribution (after correcting for the motion of the Earth) 7-9 kpc from the halo center, and bin results by the distance to the nearest larger halo.  The results are shown in figure \ref{f:vdf}.  While there is a significant increase in the high-energy velocity distribution at 8kpc for Milky Way-sized halos about to cross the virial radius of a larger halo, this effect is gone when the larger halo is $\ge 2\Rvir$ away.  As such, the presence of M31 should not impact dark matter detection experiments.

\subsection{Hostless Supernovae and GRBs}
\label{s:hostless}
\begin{figure*}
\begin{centering}
\plotgrace{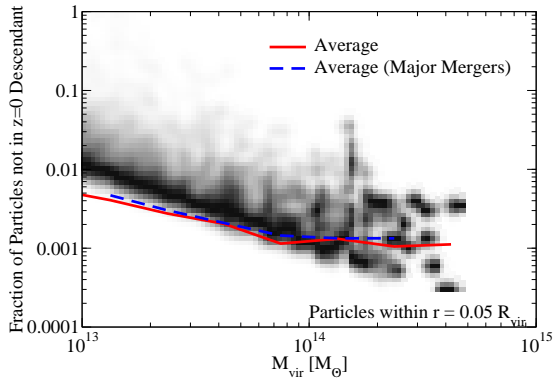}
\caption{The conditional probability distribution of particles within 0.05 $\Rvir$ of halo centers at $z=1$ which escape the halo radius by $z=0$, from the \textit{Consuelo} simulation.  This fraction is so small that many halos do not have \textit{any} particles in that radius which leave the halo; these are not shown on the density plot, but they influence the overall averages.  The fact that the average fraction is so small suggests that large-scale gravitational forces which affect halo mass accretion and mergers are unable to significantly accelerate much material out of halo centers.  For example, if more than 1\% of supernovae are undergoing hypervelocity escape, then the source of the kinetic energy will most likely be from the supernova progenitors, rather than from the halo potential well.}
\end{centering}
\label{f:statistics}
\end{figure*}

Stars are expected to form within a small radius of the center of halos \cite{Kravtsov13}, typically within 1.5-5\% of $\Rvir$.  To obtain a constraint on how many stars are expected to leave due to merging events, we can tag dark matter particles in halos within $0.05\Rvir$ at a given redshift and track how many leave their halos at later times.  Clearly, this will result in an upper bound on the escape fraction, since stars are in general much more bound than the nearby dark matter.

Nonetheless, this upper bound is an interesting one.  We trace particles in the Consuelo simulation between $z\sim1$ and $z=0$ to see which of them actually remain in the halo with which they were associated at $z=1$.  For every progenitor halo at $z=1$, we can find the descendant halo at $z=0$ which receives the largest fraction of the progenitor's particles.  The progenitor particles which do not end up in the descendant may then be assumed to have escaped sometime between $z=1$ and $z=0$.

For particles which were initially within $0.05\Rvir$, figure \ref{f:statistics} shows that 0.5\% or less can escape the halo's descendant completely by $z=0$.  In terms of hostless supernovae and GRBs, this would imply that it is extremely difficult for central dark matter particles to acquire enough energy to leave their hosts.  Thus, depending on the host halo mass, the fraction of stars (i.e., supernovae and GRB progenitors) which receive enough of a kick in halo-halo mergers to leave the halo is minimal.  Confirmed hostless high-energy events above this fraction must therefore be due to progenitor-linked energy sources, rather than long-range gravitational sources.

\subsection{The Effect of Unbound Particles on the Baryon Fraction}

\label{s:baryons}

\begin{figure}
\begin{center}
\vspace{-5ex}
\plotlargegrace{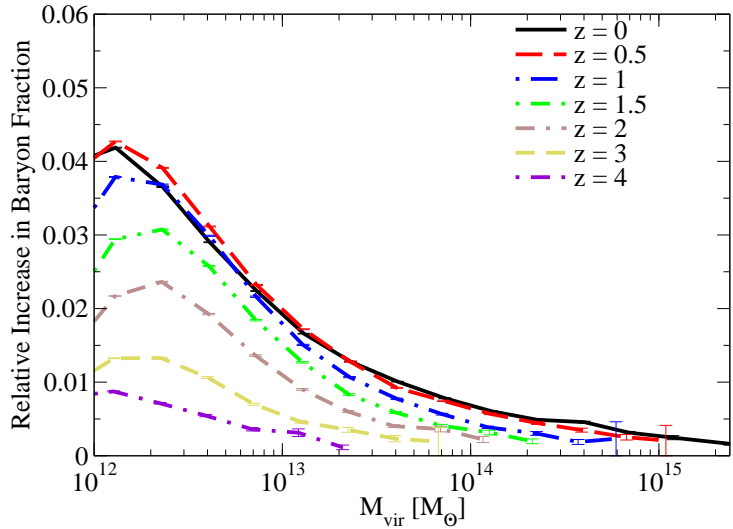}
\vspace{-5ex}
\end{center}
\caption{In satellite mergers, high-energy incoming dark matter particles will tend to escape halos, whereas the more tightly-bound stellar material will tend to remain.  This will enhance the baryon fraction over the cosmic average; this figure shows the relative increase in the baryon fraction as a function of halo mass and redshift due to this effect.  For small halos, baryonic feedback effects (such as supernovae and stellar winds) overwhelm the importance of this effect and decrease the baryon fraction below the cosmic average.  For large halos, the effect is too small to impact current cosmological surveys.  Errors shown are jackknife uncertainties in the average across halos at a given mass and redshift.}
\label{f:baryon_f}
\end{figure}

As mentioned in the introduction, high-energy dark matter particles can freely pass through dark matter halos, whereas more tightly-bound stars will tend to remain inside the halo (see \S\ref{s:full_sims}).  This effect is one of many which will influence the baryon fraction, including galaxy outflows, shocks, and angular momentum exchange.  Indeed, galaxy outflows have a much stronger impact than any other mechanism for lower-mass halos ($M_h<10^{13}\Msun$), as the energy transfer in feedback effects can be comparable to the binding energy of the baryons.  However, in high-mass clusters ($M_h>10^{14}\Msun$), the binding energy is large enough that it is very difficult for baryons to escape the halo once they are accreted.

To estimate the effects due to the different binding energies of dark matter and stars, we evaluate a toy model on all particles in the Consuelo simulation.  We assume that dark matter particles entering a halo for the first time will be accompanied by baryons according to the cosmic fraction $f_b$ (0.16, for the Consuelo cosmology).  Some fraction $f_\ast$ of the baryons will then be converted into stars; in this analysis, we use the results of \cite{Behroozi12}, which observationally constrain $f_\ast$ as a function of halo mass from $z=0$ to $z=8$.  When a dark matter particle leaves, the hot gas will undergo adiabatic expansion in the reduced potential well, expelling a fractional mass $f_b (1-f_\ast)$ of baryons.\footnote{This fraction is valid for adiabatic simulations, in which the gas fraction is roughly independent of radius \cite{Kravtsov05}.  However, this is not necessarily the case in simulations with radiative cooling; so we leave a more detailed analysis for future work.}  Any stars formed will remain in the halo, resulting in a net increase in the baryon fraction.

This model serves to put a lower limit on the baryon fraction increase from this effect.  The presence of cold gas will result in even more baryons staying behind when dark matter particles leave.  However, cold gas is limited to the same lower-mass halos which are strongly influenced by galaxy outflows---a much larger uncertainty than the effect examined here.

Predictions for the relative increase in the baryon fraction as a function of halo mass and redshift are shown in figure \ref{f:baryon_f}.  There is a strong trend towards larger increases in the baryon fraction for smaller halos, which is mainly due to the much stellar mass fraction in such halos \cite{Behroozi12}.  The relative change in the baryon fraction also increases at later redshifts; this is likely related to the smaller fraction of unbound particles at high redshifts (figure \ref{f:radial}) as well as the increasing ease of escape at later times due to the acceleration of the universe (figure \ref{f:sims}).

Overall, at $z=0$, the relative increase in the baryon fraction is between 0.5-4\%.  For very massive halos, the effect is completely negligible, as the average stellar mass to dark matter mass (including satellites and intracluster light) is less than 1\%.  Consequently, it is unlikely that this effect will influence precision calibrations of the dark matter halo mass function.  For smaller halos, the effect seems to balance the effect of angular momentum exchange, which would otherwise reduce the baryon fraction by $\sim$3\% \cite{Kravtsov05}; yet, as mentioned above, galactic outflows are likely to be significantly more important for these halos.

\section{Conclusions}

\label{s:conclusions}

We have studied the properties of unbound dark matter particles, defined as particles which can travel arbitrarily far from their halo in the far future.  Our main findings are as follows:
\begin{enumerate}
\item On average, at $z=0$, 2--6\% of particles in halos are unbound (i.e., can escape to infinity; \S \ref{s:radial}).
\item The fraction of unbound particles in halos is a very strong function of redshift, and approximately a linear function of distance from the halo center (\S \ref{s:radial}).
\item Total kinetic and potential energies cannot be used to predict which particles will become unbound.  Moreover, total kinetic and potential energies are strongly non-conserved at the single-particle level in simulations.  The level of non-conservation is independent of simulation resolution (\S \ref{s:full_sims}).  
\item The presence of Andromeda does not impact the velocity distribution of dark matter in the Milky Way at the galactocentric radius (\S \ref{s:dm_detect}).
\item Halo mergers cannot be used to explain a hostless supernovae or gamma ray burst fraction which exceeds the order of 1\% (\S \ref{s:hostless}). 
\item The baryon fraction in massive ($10^{14}$--$10^{15}\Msun$) halos is negligibly boosted ($0.5-1\%$) on account of dark matter particles which leave after entering the halo (\S \ref{s:baryons}).
\end{enumerate}

\acknowledgments

We are especially indebted to Matthew R.\ Becker for supplying the main simulation used in this work, as well as to Douglas Rudd and the University of Chicago Research Computing Center for support in running it.  We also very much appreciate the many improvements to this paper made at the suggestion of our referee, Andrey Kravtsov.  We thank Michael Busha, who ran the Consuelo and Esmeralda simulations used here, on the Orange cluster at SLAC; these were run as a part of the LasDamas collaboration simulations.  We also appreciate Joel Primack and Anatoly Klypin for providing access to the Bolshoi simulation.  Finally, we would like to thank Dusan Keres and Tom Abel for useful discussions on baryon fractions in hydrodynamical simulations, as well as Paul Schechter for other helpful comments.  This work received support from Stanford University and the U.S. Department of Energy under contract number DE-AC02-76SF00515.

\appendix

\section{The Boundedness of a Test Particle in a Spherical Potential for Arbitrary Trajectories}
\label{s:boundedness}

Knowing that angular momentum ($r v_t$, where $v_t$ is tangential velocity) will be conserved, we can write down the radial velocity $v_r$ of a particle relative to a point mass as a function of its radius $r$:
\begin{equation}
v_r(r) = \sqrt{v_0^2 + 2GM\left(\frac{1}{r}-\frac{1}{r_0}\right) -qH^2(r^2-r_0^2) - \left(\frac{r_0}{r}v_{t,0}\right)^2}
\end{equation}
where $v_{0}$ is the original velocity of the particle and $v_{t,0}$ is the original tangential component.  

Some further simplification is possible if we note that the following is a conserved quantity for a test particle:
\begin{equation}
\label{e:e_def}
E_{h} = KE + PE + \frac{1}{2}qH^2r^2
\end{equation}
Thus, we can write\footnote{As before, this applies equally well to a spherical matter distribution which is contained within $r_0$}
\begin{equation}
v_r(r) = \sqrt{\frac{2GM}{r} - 2E_h - qH^2r^2 - \left(\frac{r_0}{r}v_{t,0}\right)^2}
\end{equation}

If the particle is unbound, then $v_r(r>0)=0$ can have at most one solution at $r \le r_0$; a bound particle will have either a solution for $r>r_0$ or, in the case of a circular orbit, $r=r_0$ will be a solution and also a local maximum of $v_r^2$.  One option for distinguishing these cases is to search for the zeros of the following quartic equation:
\begin{equation}
\label{e:zeros}
-qH^2 r^4 - 2E_h r^2+2GMr - r_0^2v_{t,0}^2 = 0
\end{equation}
A somewhat simpler option is to search for the minimum of $v_r^2$ and check to see if it is negative; the locations of the extrema of $v_r^2$ are given by:
\begin{equation}
-2qH^2r^4 -2GMr + 2r_0^2v_{t,0}^2 = 0
\label{e:extrema}
\end{equation}
Before solving this latter equation, recall that a particle on the threshold of boundedness will be one where $v_r^2$ just barely reaches zero; thus, the location of the minimum of $v_r^2$ and the location of the root will coincide at the turnaround radius.  Hence, we can substitute Eq.\ \eqref{e:extrema} into Eq.\ \eqref{e:zeros} to reduce the equation to a quadratic:
\begin{equation}
\label{e:quad}
E_h r^2+\frac{3}{2}GMr - r_0^2v_{t,0}^2 = 0
\end{equation}
This gives a simple boundedness formula for $E_h$:
\begin{equation}
\label{e:eh_limit2}
E_h < \frac{-\frac{3}{2}GMr + r_0^2v_{t,0}^2}{r^2}
\end{equation}
We use Ferrari's method to approach the quartic in Eq.\ \eqref{e:extrema}.  The solution is:
\begin{eqnarray}
\label{e:r_sol}
r & = & \frac{1}{2}\left(\sqrt{y} + \sqrt{-y+\frac{2GM}{-qH^2\sqrt{y}}}\right)\\
y & = & 2\left(\sqrt[3]{R} + \frac{K}{\sqrt[3]{R}}\right)\\
R & = & J + \sqrt{J^2-K^3}\\
J & = & \frac{G^2M^2}{16q^2H^4} = \frac{r_e^6}{16}\\
K & = & \frac{r_0^2v_{t,0}^2}{-3qH^2}
\end{eqnarray}
Of course, there are three more solutions to the quartic; the full solution set is
\begin{equation}
r = \frac{1}{2}\left(\pm_1\sqrt{y} \pm_2 \sqrt{-y\pm_1\frac{2GM}{-qH^2\sqrt{y}}}\right)\\
\end{equation}
where the $\pm_1$ correspond to identical signs, and the $\pm_2$ is independent.  However, the solution in Eq.\ \eqref{e:r_sol} is the only one which corresponds to $r = r_e$ (the equivalence radius) for radial motion ($v_{t,0} = 0$).

These equations have been verified as correct in gravitational simulations of a particle escaping a halo.  Admittedly, they are quite inelegant, and in some cases, a simpler expression may be desired.  A useful \textit{approximation} is to require that angular momentum be conserved at the equivalence radius; this results in the following condition for $v_{t,0}>0$:
\begin{equation}
\label{e:eh_approx}
E_h \lesssim -\frac{3\sqrt[3]{q}}{2}(HG\mvir)^\frac{2}{3} + \frac{1}{2}r_0^2v_{t,0}^2 \left(\frac{-qH^2}{G\mvir}\right)^\frac{2}{3}
\end{equation}
For a particle with purely tangential velocity at the virial radius of a halo, this gives the correct bound for $E_h$ to within about 1\%.

\section{Kinetic Energy Thresholds in an Expanding Cosmology}

\label{s:num_sim}

As discussed in \S\ref{s:full_sims}, kinetic energy thresholds are \textit{not} a good proxy for determining which dark matter particles are bound or unbound.  However, it is still interesting to consider how different the energy thresholds would be for non-dark matter particles launched from inside halos---e.g., hypervelocity stars or galaxy outflows.

In order to define the appropriate kinetic energy cuts, we perform a series of numerical simulations for particle trajectories in an expanding Universe.  The evolution of the scale factor with time is given by the standard equation in a flat universe:
\begin{equation}
a(t) = \sqrt[3]{\frac{\Omega_m}{\Omega_\Lambda}\sinh^2\left(\frac{3}{2}H_0 t \sqrt{\Omega_\Lambda}\right)}
 \end{equation}
 We adopt $\Omega_m = 0.25$, $\Omega_\Lambda = 0.75$, and $H_0 = 70$ km s$^{-1}$ Mpc$^{-1}$ for consistency with the cosmological simulation we have used.
 
\begin{figure}
\plotgrace{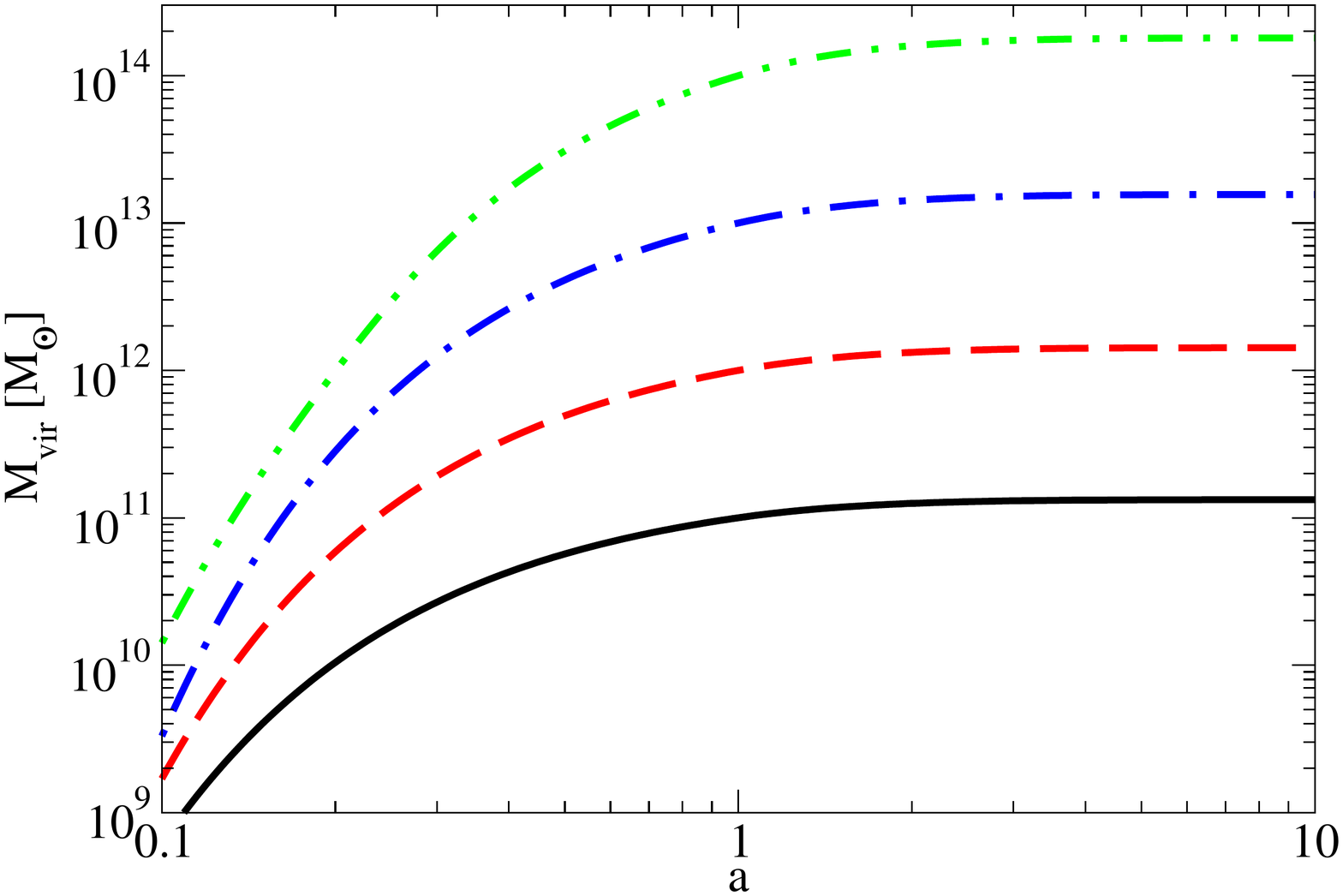}\plotgrace{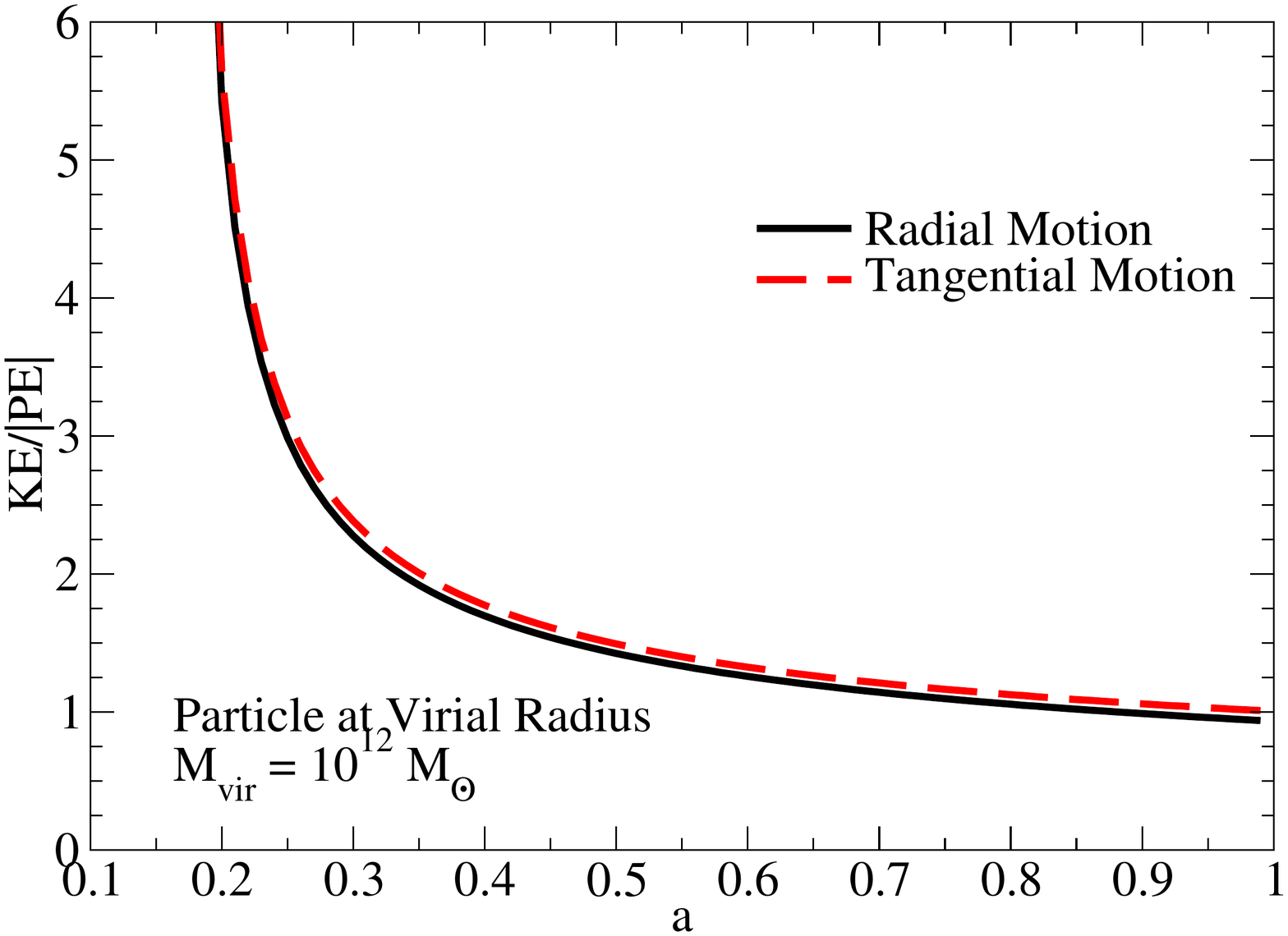}\\[-3ex]
\plotgrace{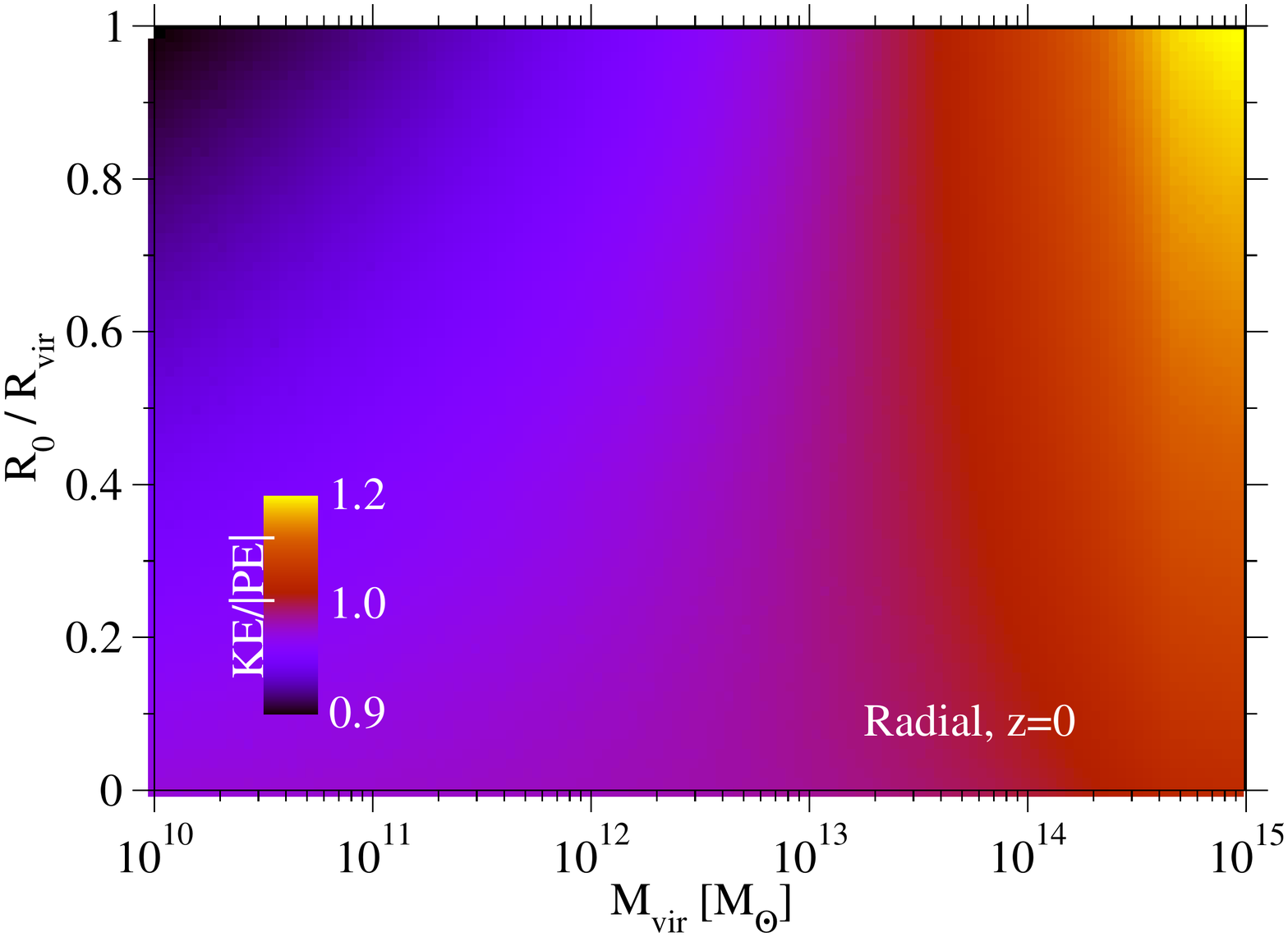}\plotgrace{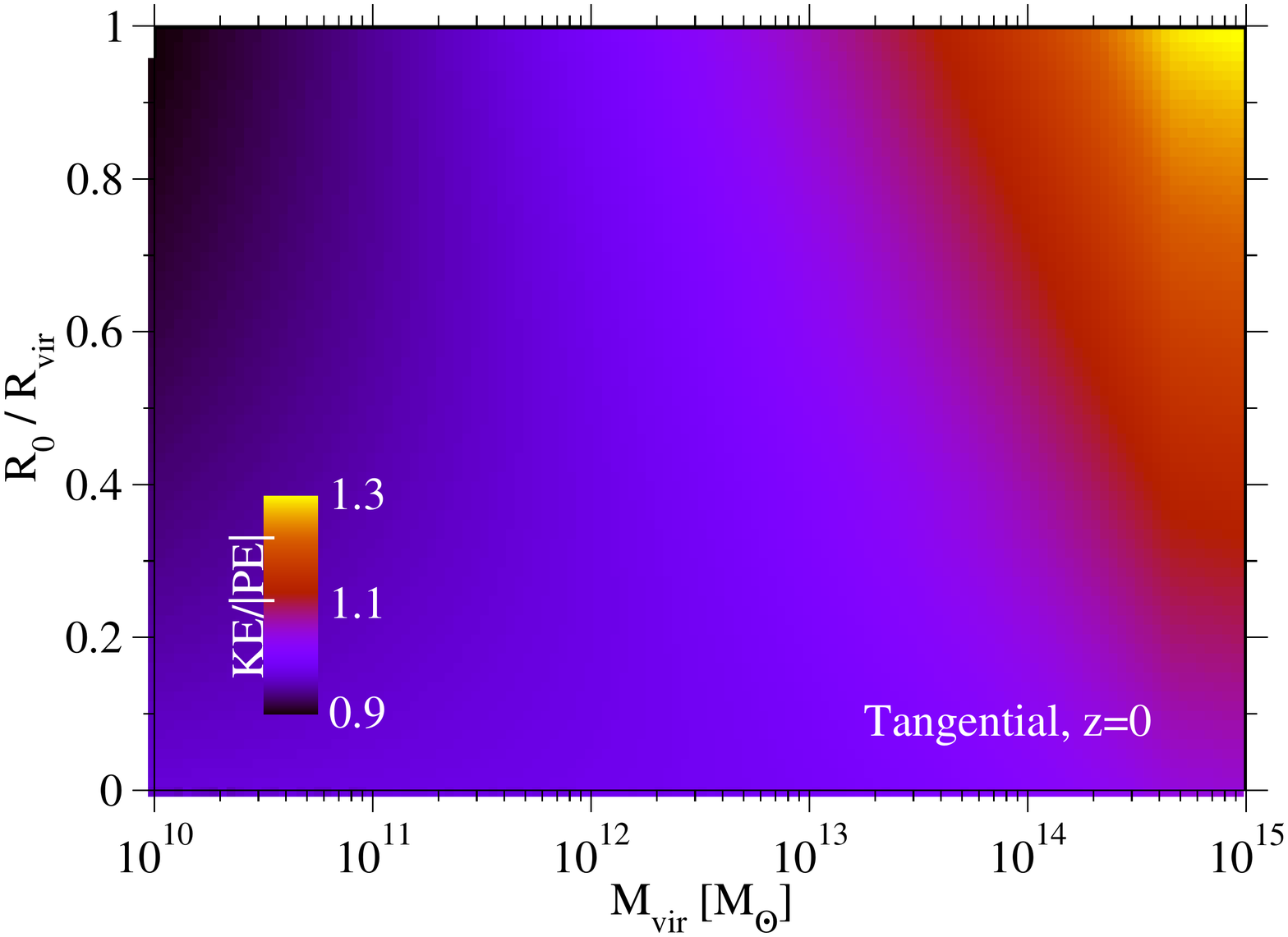}\\[-3ex]
\caption{\textbf{Top-left} panel: The mass accretion history and future for $10^{11}$, $10^{12}$, $10^{13}$, and $10^{14}\Msun$ halos at $z=0$ as given by Eqns.\ \eqref{e:mah}-\eqref{e:mah_a}.  Beyond a scale factor of $a=3$, further accretion is practically negligible.  \textbf{Top-right} panel: the minimum kinetic energy necessary to escape for a $10^{12}\Msun$ halo as a function of redshift.  The Universe decelerates at high redshifts, making it much more difficult for particles to escape to infinity.  For ease of comparison, the kinetic energies are scaled to the initial potential energy of the particle, counting only particles within the host halo's virial radius.  The \textbf{bottom-left} panel shows the minimum kinetic energy as a function of halo mass and radius for radial particle trajectories at $z=0$.  Finally, the \textbf{bottom-right} panel shows minimum kinetic energies for tangential trajectories, also at $z=0$.}
\label{f:sims}
\end{figure}

As mentioned at the end of \S \ref{s:analytic}, halos are always in the process of accreting matter, so the gravitating mass affecting an escaping particle will be changing as well.  For these numerical simulations, we model halos as having spherical NFW \cite{NFW97} profiles with concentrations given by \cite{Prada11}.\footnote{We do not find evidence for a steeper fall-off than $r^{-3}$ until $\sim 5\Rvir$ in our simulations at $z=0$.}  We truncate the halo mass profile at the radius where the matter density matches the critical density.  Halo mass accretion histories are accounted for by a slight modification to the standard formula from \cite{Wechsler02}:
\begin{equation}
\label{e:mah}
\mvir(a) = M_0 \exp(z f(M_0, a))
\end{equation}
We allow the extra dependence on redshift on account of the fact that the far future mass accretion rates are somewhat less than those predicted by the mass accretion histories in \cite{Wechsler02}.  Specifically, based on the mass accretion histories to $a=100$ in \cite{Busha05} and the mass accretion histories for $a\le 1$ from the merger trees in \cite{BehrooziTree}, we adopt the following functional form for $f(M_0, a)$:
\begin{eqnarray}
f(M_0, a) & = & g(a)\left[- 0.122\log_{10}\left(\frac{M_0}{10^{6.47}\Msun}\right)\right.\nonumber\\
	&&  \left. -0.328\exp\left(\log_{10}\frac{M_0}{10^{14.24}\Msun}\right)\right]\\
g(a) & = & \left\{
\begin{array}{rl} 1 & \textrm{if }a\le 1\\
0.5+0.5\left(\frac{2}{1+a}\right)^{1.3} & \textrm{if }a>1\end{array} \label{e:mah_a}
\right.
\end{eqnarray}
An example of the mass accretion history (and future) for $10^{11}$ to $10^{14} \Msun$ halos at $z=0$ is shown in the top-left panel of figure \ref{f:sims}.

With these assumptions, we can evaluate the trajectory of a particle released in a cosmological environment as a function of initial radius, velocity, host halo mass, and redshift.  If a particle crosses the equivalence radius (Eq.\ \eqref{e:r_e}), it will escape to infinity, which provides a straightforward test for boundedness.\footnote{Note that the equivalence radius always shrinks with increasing time, eventually asymptoting to a fixed multiple of the virial radius.}

We show results from the numerical simulations in figure \ref{f:sims}.  Notably, at $z=0$, the kinetic energy thresholds are within 30\% of the expected Newtonian values.  However, at high redshifts, the energy required for escape increases dramatically due to the deceleration of the Universe.  By $z=2$, the kinetic energy required for escape from a $10^{12}\Msun$ halo is already twice the Newtonian-expected value, and by $z=4$, it is six times as much.  This effect is a likely explanation for the reduced fraction of unbound dark matter particles seen at high redshifts in \S \ref{s:radial}.

\section{The Effect of Substructure on Escaping Particles}

\label{s:substructure}
Consider a host halo with mass $M_{host}$ and virial radius $R_{host}$.  In such a halo, the particle escape velocity $V$ will be proportional to $\sqrt{\frac{GM_{host}}{R_{host}}}$.  An escaping particle which interacts with a subhalo with mass $M_{sub}$ and radius $R_{sub}$ will experience a change in velocity ($\Delta V$) equal to the average acceleration multiplied by the crossing time.  The crossing time will be proportional to $R_{sub} / V$, which gives
\begin{equation}
\frac{\Delta V}{V} = \frac{|a|t}{V} \propto \frac{GM_{sub}}{R_{sub}^2} \cdot \frac{R_{sub}}{V} \cdot \frac{1}{V} = \frac{M_{sub}}{R_{sub}} \frac{R_{host}}{M_{host}}
\end{equation}
As radius is proportional to $M^{1/3}$ in both cases, we conclude
\begin{equation}
\label{e:dv}
\frac{\Delta V}{V} \propto \left(\frac{M_{sub}}{M_{host}}\right)^{2/3}
\end{equation}
The largest subhalo typically found in host halos is between 1/20 and 1/15 the size of the host halo \cite{Zheng07}.  Eq.\ \eqref{e:dv} would suggest a $\sim$ 15\% change (within a factor of a few) in the velocity of the escaping particle.  However, this velocity change is primarily a change in the \textit{direction} of the particle, rather than its kinetic energy.

To see this latter point, note that if the gravitational potential of the subhalo is time-independent, it must be energy-conserving.  Thus, as long as the particle was initially far from the subhalo, passing through the subhalo will not affect its total kinetic plus potential energy \textit{unless the subhalo itself changes during the time that the particle is passing through it}.  From this, it should be clear that the kinetic energy change will be small.  However, to be quantitative about this statement, we note that the primary causes of large changes to the subhalo will be major mergers (rare) as well as tidal disruption.  We may estimate the tidal disruption timescale as the dynamical time of the host halo ($\propto\sqrt{\frac{R_{host}^3}{GM_{host}}}$).  The change in energy $\Delta E$ from the subhalo will then be proportional to the change in the potential of the subhalo during the particle crossing time ($\sim \frac{G\Delta M_{sub}}{R_{sub}} \propto \frac{GM_{sub}}{R_{sub}}  \cdot \frac{\Delta t}{t_{dyn}}$), so we find that:
\begin{eqnarray}
\label{e:de}
\frac{\Delta E}{E} & \propto & \frac{\Delta E}{V^2} \propto \frac{G M_{sub}}{R_{sub}} \cdot \frac{R_{sub}}{V} \cdot \sqrt{\frac{GM_{host}}{R_{host}^3}} \cdot \frac{1}{V^2}\\
& \propto & \frac{M_{sub}}{M_{host}}
\end{eqnarray}
Thus, even for the largest typical subhalo, the resulting kinetic energy change is within a factor of a few of the subhalo mass ratio ($5-6\%$).  This, even in combination with the direction change (which will result in a small change in the energy threshold for escape), will change the eventual fate (bound or unbound) of only a tiny fraction of particles.

This argument does not work for dark matter particles because the primary \textit{source} of high-energy dark matter particles in halos is incoming substructure.  Hence, the substructure crossing time used here is a severe underestimate for the relevant dark matter particles.  Nonetheless, the approach used here is reasonable for particles ejected through other means from the central galaxy of a halo.

One last feature which is worth noting: the positive sign of the energy change in Eq.\ \eqref{e:de} is always correct within the virial radius of the host.  An escaping particle encountering a satellite will experience less gravitational force on the way out than on the way in because the satellite is becoming more tidally dispersed with time.  Thus, the total acceleration will typically be larger than the total deceleration, and the net effect of the encounter will be a slight energy gain by the escaping particle.  It is only outside the virial radius of the host halo that the sign can become negative, because there it is possible for other halos to grow in mass with time.

\newpage 

%
%  These Macros are taken from the AAS TeX macro package version 5.2
%  and are compatible with the macros in the A&A document class
%  version 7.0
%  Include this file in your LaTeX source only if you are not using
%  the AAS TeX macro package or the A&A document class and need to
%  resolve the macro definitions in the TeX/BibTeX entries returned by
%  the ADS abstract service.
%
%  If you plan not to use this file to resolve the journal macros
%  rather than the whole AAS TeX macro package, you should save the
%  file as ``aas_macros.sty'' and then include it in your LaTeX paper
%  by using a construct such as:
%	\documentstyle[11pt,aas_macros]{article}
%
%  For more information on the AASTeX and A&A packages, please see:
%	http://aastex.aas.org
%       ftp://ftp.edpsciences.org/pub/aa/readme.html
%  For more information about ADS abstract server, please see:
%	http://adsabs.harvard.edu/ads_abstracts.html
%

% Abbreviations for journals.  The object here is to provide authors
% with convenient shorthands for the most "popular" (often-cited)
% journals; the author can use these markup tags without being concerned
% about the exact form of the journal abbreviation, or its formatting.
% It is up to the keeper of the macros to make sure the macros expand
% to the proper text.  If macro package writers agree to all use the
% same TeX command name, authors only have to remember one thing, and
% the style file will take care of editorial preferences.  This also
% applies when a single journal decides to revamp its abbreviating
% scheme, as happened with the ApJ (Abt 1991).

\def\ref@jnl#1{{\rm#1}}

\def\aj{\ref@jnl{AJ}}                   % Astronomical Journal
\def\actaa{\ref@jnl{Acta Astron.}}      % Acta Astronomica
\def\araa{\ref@jnl{ARA\&A}}             % Annual Review of Astron and Astrophys
\def\apj{\ref@jnl{ApJ}}                 % Astrophysical Journal
\def\apjl{\ref@jnl{ApJ}}                % Astrophysical Journal, Letters
\def\apjs{\ref@jnl{ApJS}}               % Astrophysical Journal, Supplement
\def\ao{\ref@jnl{Appl.~Opt.}}           % Applied Optics
\def\apss{\ref@jnl{Ap\&SS}}             % Astrophysics and Space Science
\def\aap{\ref@jnl{A\&A}}                % Astronomy and Astrophysics
\def\aapr{\ref@jnl{A\&A~Rev.}}          % Astronomy and Astrophysics Reviews
\def\aaps{\ref@jnl{A\&AS}}              % Astronomy and Astrophysics, Supplement
\def\azh{\ref@jnl{AZh}}                 % Astronomicheskii Zhurnal
\def\baas{\ref@jnl{BAAS}}               % Bulletin of the AAS
\def\bac{\ref@jnl{Bull. astr. Inst. Czechosl.}}
                % Bulletin of the Astronomical Institutes of Czechoslovakia 
\def\caa{\ref@jnl{Chinese Astron. Astrophys.}}
                % Chinese Astronomy and Astrophysics
\def\cjaa{\ref@jnl{Chinese J. Astron. Astrophys.}}
                % Chinese Journal of Astronomy and Astrophysics
\def\icarus{\ref@jnl{Icarus}}           % Icarus
\def\jcap{\ref@jnl{J. Cosmology Astropart. Phys.}}
                % Journal of Cosmology and Astroparticle Physics
\def\jrasc{\ref@jnl{JRASC}}             % Journal of the RAS of Canada
\def\memras{\ref@jnl{MmRAS}}            % Memoirs of the RAS
\def\mnras{\ref@jnl{MNRAS}}             % Monthly Notices of the RAS
\def\na{\ref@jnl{New A}}                % New Astronomy
\def\nar{\ref@jnl{New A Rev.}}          % New Astronomy Review
\def\pra{\ref@jnl{Phys.~Rev.~A}}        % Physical Review A: General Physics
\def\prb{\ref@jnl{Phys.~Rev.~B}}        % Physical Review B: Solid State
\def\prc{\ref@jnl{Phys.~Rev.~C}}        % Physical Review C
\def\prd{\ref@jnl{Phys.~Rev.~D}}        % Physical Review D
\def\pre{\ref@jnl{Phys.~Rev.~E}}        % Physical Review E
\def\prl{\ref@jnl{Phys.~Rev.~Lett.}}    % Physical Review Letters
\def\pasa{\ref@jnl{PASA}}               % Publications of the Astron. Soc. of Australia
\def\pasp{\ref@jnl{PASP}}               % Publications of the ASP
\def\pasj{\ref@jnl{PASJ}}               % Publications of the ASJ
\def\rmxaa{\ref@jnl{Rev. Mexicana Astron. Astrofis.}}%
                % Revista Mexicana de Astronomia y Astrofisica
\def\qjras{\ref@jnl{QJRAS}}             % Quarterly Journal of the RAS
\def\skytel{\ref@jnl{S\&T}}             % Sky and Telescope
\def\solphys{\ref@jnl{Sol.~Phys.}}      % Solar Physics
\def\sovast{\ref@jnl{Soviet~Ast.}}      % Soviet Astronomy
\def\ssr{\ref@jnl{Space~Sci.~Rev.}}     % Space Science Reviews
\def\zap{\ref@jnl{ZAp}}                 % Zeitschrift fuer Astrophysik
\def\nat{\ref@jnl{Nature}}              % Nature
\def\iaucirc{\ref@jnl{IAU~Circ.}}       % IAU Cirulars
\def\aplett{\ref@jnl{Astrophys.~Lett.}} % Astrophysics Letters
\def\apspr{\ref@jnl{Astrophys.~Space~Phys.~Res.}}
                % Astrophysics Space Physics Research
\def\bain{\ref@jnl{Bull.~Astron.~Inst.~Netherlands}} 
                % Bulletin Astronomical Institute of the Netherlands
\def\fcp{\ref@jnl{Fund.~Cosmic~Phys.}}  % Fundamental Cosmic Physics
\def\gca{\ref@jnl{Geochim.~Cosmochim.~Acta}}   % Geochimica Cosmochimica Acta
\def\grl{\ref@jnl{Geophys.~Res.~Lett.}} % Geophysics Research Letters
\def\jcp{\ref@jnl{J.~Chem.~Phys.}}      % Journal of Chemical Physics
\def\jgr{\ref@jnl{J.~Geophys.~Res.}}    % Journal of Geophysics Research
\def\jqsrt{\ref@jnl{J.~Quant.~Spec.~Radiat.~Transf.}}
                % Journal of Quantitiative Spectroscopy and Radiative Transfer
\def\memsai{\ref@jnl{Mem.~Soc.~Astron.~Italiana}}
                % Mem. Societa Astronomica Italiana
\def\nphysa{\ref@jnl{Nucl.~Phys.~A}}   % Nuclear Physics A
\def\physrep{\ref@jnl{Phys.~Rep.}}   % Physics Reports
\def\physscr{\ref@jnl{Phys.~Scr}}   % Physica Scripta
\def\planss{\ref@jnl{Planet.~Space~Sci.}}   % Planetary Space Science
\def\procspie{\ref@jnl{Proc.~SPIE}}   % Proceedings of the SPIE

\let\astap=\aap
\let\apjlett=\apjl
\let\apjsupp=\apjs
\let\applopt=\ao

\bibliography{unbound-jcap}

\end{document}